\documentclass[journal=aamick,manuscript=article]{achemso}

\usepackage[version=3]{mhchem} 
\usepackage{color}
\usepackage{placeins}

\author{Gustav Eriksson}
\email{gustav.eriksson@chalmers.se}
\affiliation[Chalmers1]{Department of Chemistry and Chemical Engineering, Chalmers University of Technology Gothenburg 41296, Sweden}
\author{Matteo De Tullio}
\affiliation[Rouen]{Universit{\'e} Rouen Normandie, INSA Rouen Normandie, CNRS, GPM UMR 6634, F-76000 Rouen, France}
\author{Francesco Carnovale}
\affiliation[UniTN]
{Department of Physics, University of Trento, Italy}
\alsoaffiliation[TIFPA]
{Trento Institute for Fundamental Physics and Applications (TIFPA), National Institute for Nuclear Physics (INFN), Italy}
\alsoaffiliation[ECT]
{European Centre for Theoretical Studies in Nuclear Physics and Related Areas (ECT*), Fondazione Bruno Kessler (FBK), Trento, Italy}
\author{Giovanni Novi Inverardi}
\affiliation[UniTN]
{Department of Physics, University of Trento, Italy}
\alsoaffiliation[TIFPA]
{Trento Institute for Fundamental Physics and Applications (TIFPA), National Institute for Nuclear Physics (INFN), Italy}
\alsoaffiliation[ECT]
{European Centre for Theoretical Studies in Nuclear Physics and Related Areas (ECT*), Fondazione Bruno Kessler (FBK), Trento, Italy}
\author{Tommaso Morresi}
\email{morresi@ectstar.eu}
\affiliation[ECT]
{European Centre for Theoretical Studies in Nuclear Physics and Related Areas (ECT*), Fondazione Bruno Kessler (FBK), Trento, Italy}
\alsoaffiliation[TIFPA]
{Trento Institute for Fundamental Physics and Applications (TIFPA), National Institute for Nuclear Physics (INFN), Italy}
\author{Jonathan Houard}
\alsoaffiliation[Rouen]
{Universit{\'e} Rouen Normandie, INSA Rouen Normandie, CNRS, GPM UMR 6634, F-76000 Rouen, France}
\author{Marc Ropitaux}
\affiliation[nuova]{Universit{\'e} Rouen Normandie, GLYCOMEV UR4358, SFR Normandie V{\'e}g{\'e}tal FED 4277, Innovation Chimie Carnot, IRIB, F-76000 Rouen, France}
\author{Ivan Blum}
\affiliation[Rouen]
{Universit{\'e} Rouen Normandie, INSA Rouen Normandie, CNRS, GPM UMR 6634, F-76000 Rouen, France}
\author{Emmanuel Cadel}
\affiliation[Rouen]
{Universit{\'e} Rouen Normandie, INSA Rouen Normandie, CNRS, GPM UMR 6634, F-76000 Rouen, France}
\author{Gianluca Lattanzi}
\affiliation[UniTN]
{Department of Physics, University of Trento, Italy}
\alsoaffiliation[TIFPA]
{Trento Institute for Fundamental Physics and Applications (TIFPA), National Institute for Nuclear Physics (INFN), Italy}
\author{Mattias Thuvander}
\affiliation[Chalmers2]
{Department of Physics, Chalmers University of Technology, Gothenburg 41296, Sweden}
\author{Martin Andersson}
\affiliation[Chalmers1]{Department of Chemistry and Chemical Engineering, Chalmers University of Technology Gothenburg 41296, Sweden}
\author{Mats Hulander}
\affiliation[Chalmers1]{Department of Chemistry and Chemical Engineering, Chalmers University of Technology Gothenburg 41296, Sweden}
\author{Simone Taioli}
\email{taioli@ectstar.eu}
\affiliation[ECT]
{European Centre for Theoretical Studies in Nuclear Physics and Related Areas (ECT*), Fondazione Bruno Kessler (FBK), Trento, Italy}
\alsoaffiliation[TIFPA]
{Trento Institute for Fundamental Physics and Applications (TIFPA), National Institute for Nuclear Physics (INFN), Italy}
\author{Angela Vella}
\email{angela.vella@univ-rouen.fr}
\affiliation[Rouen]{Universit{\'e} Rouen Normandie, INSA Rouen Normandie, CNRS, GPM UMR 6634, F-76000 Rouen, France}
\title[]
  {Role of defects in atom probe analysis of sol-gel silica
   }


\begin{document}







\newpage
\begin{abstract}
Silicon dioxide is a suitable material to encapsulate proteins at room temperature so that they can be analysed at the atomic level using laser-assisted atom probe tomography (La-APT).
To achieve this goal, in this study we show that UV and deep-UV lasers can achieve a high success rate in La-APT of silica in terms of chemical resolution and three-dimensional image volume, with both lasers providing comparable results. Since the La-APT analyses are driven by photon absorption, in order to understand the mechanisms behind the enhanced absorption of UV light, we performed density functional theory calculations to model the electronic and optical properties of amorphous silica matrices generated using a Monte Carlo approach to structural optimisation.
In particular, we have investigated the role of various defects introduced during sample preparation, such as substitutional and interstitial carbon, sodium and gallium ions, and hydrogen.
Our results show that the presence of defects increases the absorption of silica in the UV and deep-UV range and thus improves the La-APT capabilities of the material. However, due to the low density of free charge carriers resulting from the absorption of laser energy by defects, deviations from the nominal chemical composition and suboptimal chemical resolution may occur, potentially limiting the optimal acquisition of APT mass spectra.
\end{abstract}

\section{Introduction}

Silicon dioxide (SiO$_2$), commonly known as silica, is the most abundant element in the earth's crust \cite{florke2000silica, wedepohl1995composition}. It occurs in nature in various forms, e.g. as sand, minerals and quartz. At the molecular level, most silica allotropes are structured as an interconnected network of silicon atoms covalently bonded in tetrahedral coordination by oxygen bridges shared by two silicon tetrahedra \cite{Douglas2006,gerber1986structure} (however, not all allotropes are tetrahedral, e.g. stishovite, which forms at very high pressure, is a polymorph in which each Si is surrounded by 6 oxygen atoms).\\
\indent 
In amorphous silicon dioxide, this local, short-range order does not continue into a longer-range order, but forms an amorphous material with characteristic properties such as transparency to visible light, brittleness and chemical inertness.\\
\indent 
Silica has a long history of being processed and utilized, especially in glass manufacturing, and has also been synthesised using different methods. One commonly used method is the Stöber process, in which a silicate precursor is hydrolysed under either alkaline (pH $\simeq$ 11-12) or basic conditions using ammonia as a catalyst and then condensed to form the silica network \cite{stober1968controlled}. Another method is the sol-gel process, in which a precursor solution, such as sodium silicate with a high pH value, forms a macroscopic gel by polymerisation through condensation of the silicate monomers when the pH value is lowered.
Sol-gel silica is particularly useful for encapsulating proteins in a solid material that retains their native structure at room temperature similar to an aqueous environment \cite{noviinverardi_2023,NOVIINVERARDI20252537}. This allows proteins to be analysed at the atomic level \cite{wang2015sol}. 
For example, Immunoglobulin G (IgG) proteins embedded in silica were imaged three-dimensionally with near-atomic resolution using laser-assisted atom probe tomography (La-APT), which provided information about
their elemental composition \cite{sundell2019atom}. 
\\ 
\indent
Atom probe tomography (APT) is a powerful technique for material characterisation based on the controlled field evaporation of individual ions from a needle-shaped specimen. These ions move along the applied field and are detected by a 2D position-sensitive detector \cite{blavette1993atom,gault2021atom}. In La-APT, the process of field ion emission is triggered by ultrafast laser pulses \cite{gault2006design}. In particular, a combination of femtosecond laser pulses and static fields is used to briefly heat the sample to reduce the energy barrier for field ion evaporation \cite{vella2011field,vella2013interaction,de2025evaporation}. Time-of-flight mass spectrometry enables the chemical identification of the ions, while the position of the impact on the detector is used to reconstruct the initial position and thus generate the 3D image of the sample by back-projection \cite{blavette1993atom, gault2021atom}.
\\ 
\indent
While silica has already been used as a framework for the encapsulation of biomolecular substrates, where the native biomolecule can be preserved by minimising mechanical stresses, it poses some major challenges for APT analysis. Indeed, the very low conductivity of amorphous silica, which is an electrical insulator but whose conductivity can increase slightly under certain conditions (impurities, doping, high temperatures or radiation) represents a challenge for APT analysis, as the applied potential cannot easily be carried to the apex of the sample to obtain the electrostatic field required for field evaporation. \cite{arnoldi2014energy,arnoldi2018thermal,caplins2020algorithm}.
In this context, recent experimental and theoretical studies have shown that THz pulses with negative polarity can trigger the evaporation of cations from nanoneedle-shaped amorphous silica specimens  efficiently \cite{de2025evaporation}.
In addition, in a previous work \cite{sundell2019atom} a green laser with a wavelength of 532 nm (corresponding to a photon energy of 2.4 eV) was applied to a silica sample, resulting in a low evaporation yield and a small analysed volume due to the fracture of the sample. \\ 
\indent
The main reason for the difficulties in La-APT analysis of silica using a green laser is related to its low absorption of green light, since the band gap of silica is about 9 eV, which is higher than the photon energy of the green laser \cite{distefano1971band}. However, defects are frequently observed in sol-gel silica and can also occur during sample preparation of the nano-needle \cite{griscom1991optical,bogdanowicz2018laser}. Defects can increase the sub-band gap light absorption and thus the ion yield of La-APT analysis.\\ 
\indent 
In this work, we have performed an experimental investigation of silica samples synthesised with our sol-gel method using La-APT devices equipped with green, ultraviolet (UV) and deep-UV lasers at 515, 343 and 258 nm, combined with a computational study of the electronic and optical absorption properties of silica based on density functional theory (DFT) simulations that account for the presence of defects within the silica matrix. DFT is widely regarded as the standard computational method to study the electronic structure of solids, although more sophisticated approaches such as the GW approximation~\cite{sio2_theor,reining_2018,taioli2009electronic,10.1063/1.4716178,elena2,elena3} or the Bethe-Salpeter equation~\cite{blase_2020,elena1} provide improved quasiparticle descriptions and thus more accurate optical absorption spectra. However, their application to the class of materials investigated in this study, which are disordered and amorphous, poses significant methodological and computational challenges, making DFT a favourable compromise between accuracy and computational cost.
We show that the defects in silica most likely increase the absorption of UV and deep-UV light. This explains why we report successful La-APT analyses with high yield and large analysed volume when we use light in this frequency range, but much less often when we use green light.
Nevertheless, the experimental results also show that the density of charge carriers generated by the absorption of laser energy is not sufficient to fully correct the problem of low conductivity in La-APT analyses of insulating materials in the UV and deep-UV range. The aim of this work is in particular to determine the specific role of defects in the light absorption of silica in order to isolate their influence from other effects in APT experiments, such as that due to heating.

\section{Materials and Methods}

\subsection{Sample preparation}

\begin{figure}[!htb]
   \centering
\includegraphics[width=1\textwidth]{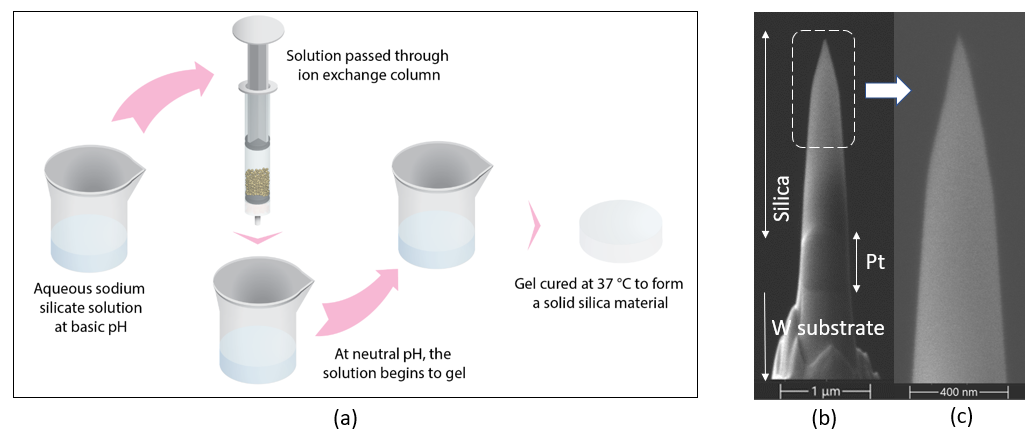}\caption{\label{fig:figure1} (a) Schematic overview of the experimental procedure for synthesising silica glass from an alkaline silicate precursor; (b-c) SEM images at different scales of an APT specimen prepared from the silica sample.}
\end{figure}

The silica samples were prepared using a sodium silicate solution (Sigma-Aldrich) as starting material, as shown in Figure \ref{fig:figure1}a. The sodium silicate solution was diluted 1:3 in water and gelation was initiated by passing the solution through a syringe filled with a Dowex 50WX ion exchange resin (Sigma-Aldrich) activated with HCl [1M] to adjust the pH from alkaline to neutral, according to the procedure described in Refs. \cite{sundell2019atom, bhatia2000aqueous}. After activation with HCl, the ion exchange resin was adjusted by passing Milli-Q water through it in small increments until an extruded pH value of around 5 was reached. The diluted sodium silicate solution was then passed through the syringe to obtain a neutral pH. A solid glass was formed by drying the silica gel at 37 °C overnight.

The specific surface area and pore size distribution of the synthesised silica samples were determined by nitrogen sorption measurements using an ASAP 2020 instrument (Micromeritics Instrument Corporation, Norcross GA). The material actually contains both micropores and mesopores. The specific surface area was assessed using the Brunauer-Emmett-Teller (BET) method \cite{brunauer1938adsorption}, the micropores with the Horvath-Kawazoe method \cite{horvath1983method} and the mesopores with the Barrett-Joyner-Halenda (BJH) sorption method \cite{barrett1951determination}. The measurements resulted in a specific surface area of 587 m$^2$/g for a representative silica sample. The micropores have an average width of 6.8 Å and the mesopores have an average size of 40 Å. Further details on the isothermal measurements can be found in the Supplementary Information (SI, see Figure S.1).\\
\indent In addition, the absorption spectrum of a thin silica sample is shown in Figure S.2 of the SI. We note that the absorption for green light is negligible and shows a slight increase at wavelengths below 400 nm and a stronger increase at wavelengths below 300 nm. \\
\indent
Finally, the APT specimens were prepared with the conventional focused ion beam scanning electron microscopy (FIB-SEM) in-situ lift-out protocol using a FIB-SEM dual beam instrument (FEI Versa 3D) with a single isotope Ga$^+$ source (69 amu) \cite{thompson_2007,blum2016atom}. An area of interest on the sample surface was identified and covered with a 2$\times$20 $\mu$m Pt protective strip 
deposited with the instrument's gas injection system (GIS) using a (methyl-cyclopentadienyl)-trimethyl platinum precursor gas. A cantilever was prepared for lift out by milling trenches with the stage tilted 30° with respect to the ion beam. The cantilever was lifted out with an Omniprobe micromanipulator and placed in segments on a flat silicon or tungsten rod. The lift-out segments were polished with the focussed ion beam (FIB) by annular milling to obtain the needle-shaped sample shown in Figure \ref{fig:figure1}b,c. We notice that the high charging effect due to the low electrical conductivity of silica reduces the image resolution.

\subsection{Atom Probe Tomography}

APT analyses were performed with two different systems: an atom probe with an energy compensator (reflectron) with a flight length of 38 cm LEAP 6000 XR (Cameca Scientific Instruments) and a linear atom probe with a flight length of 10 cm La-WATAP (from Cameca). For both systems, the measurements were carried out in an ultra-high vacuum of 10$^{-7}$ Pa.\\
\indent 
The LEAP 6000 XR operated in laser mode (wavelength 257.5 nm, $\approx$4.8 eV, deep-UV) with laser pulse energies (LPE) between 100 and 250 pJ focussed on a spot of 2 $\mu$m diameter, which corresponds to an energy density between 30 and 80 J/m$^2$. The sample temperature was set to either 35 K or 50 K. The pulse rate was set using the instrument's automatic pulse rate control so that ions could be detected up to 250 Da before the next pulse, regardless of the voltage. This meant that the pulse rate increased with increasing voltage due to shorter flight times. \\
\indent 
The La-WATAP uses a ytterbium-doped laser operating at 1030 nm. This laser generates ultrashort pulses with a duration of 350 fs at a repetition rate of 100 kHz. The laser wavelength of 1030 nm is converted into its second and third harmonics at 515 nm ($\approx$2.4 eV, green) and 343 nm ($\approx$3.6 eV, UV) using two beta-barium borate (BBO) crystals. The laser energy of the 343 nm beam can vary between 10-50 nJ per pulse and is focussed on the sample in a spot with a diameter of 30 $\mu$m, which corresponds to an energy density between 17 and 70 J/m$^2$.\\
\indent 
The APT data obtained were reconstructed with AP Suite 6.2 (Cameca Scientific Instruments, Madison WI) for LEAP analyses and GPM 3D software for La-WATAP analyses using a tip profile protocol.

\subsection{Theoretical and computational methods}

To produce defect-free amorphous silicon dioxide matrices, we assume ideal beta-cristobalite structures with a density of $0.0662$ atoms/\AA$^3$. The Monte Carlo (MC) approach developed by Wooten, Winer and Weaire \cite{www} is then used to optimise the atomic configurations. In this method, the structure of the material is represented as a network of interconnected atoms, to which we first make a series of random bond changes, all of which are accepted. This results in a randomised structure, which was then annealed at constant volume by lowering the temperature at each step and performing a series of atomic displacements to optimise the bonding topology \cite{bomc}. The criterion for accepting (rejecting) such displacements is chosen according to the Boltzmann probability distribution, where the decrease (increase) in potential energy is evaluated using a force field parameterised for SiO$_2$ \cite{keating}, similar to molecular dynamics simulations \cite{mdref1,mdref2,mdref3,mdref4}.
In this framework, we can amorphise the structures independently of their initial configurations and find minimum-energy configurations by annealing in a very efficient way. The resulting defect-free amorphous silica matrices are contained in a cubic box with periodic boundary conditions. In particular, we have generated three different sizes of simulation cells corresponding to a cubic box with an edge of $L=10.283$ \AA, $L=14.26$ \AA \ and $L=21.39$ \AA \ (see Figure S.3 of the SI for an image of one of the structures obtained).\\
\indent 
The optical properties of these systems are evaluated in the framework of linear response theory by calculating the imaginary part of the complex dielectric tensor $\epsilon^{(2)}_{\alpha,\beta}$ (where $\alpha,\beta$ denotes the tensor components with respect to a set of Cartesian axes), whose spatial average is linearly related to the absorption coefficient (modulo some weighting factors).
In our calculations, local field effects (LFE) are neglected and the expression for $\epsilon^{(2)}_{\alpha,\beta}$ is given by \cite{taioli2009electronic,taioli2024advancements}
\begin{equation}\label{epsilon}
    \epsilon^{(2)}_{\alpha,\beta} (\omega) = \frac{4 \pi e^2}{\Omega N_{\mathbf{k}}m^2} 
    \sum_{n,n'}\sum_{\mathbf{k}} \frac{\hat{\mathbf{M}}_{\alpha,\beta;n,n',\mathbf{k}}}{\left(E_{n',\mathbf{k}}-E_{n,\mathbf{k}}\right)^2} \left[ \delta \left( E_{n',\mathbf{k}}-E_{n,\mathbf{k}} + \hbar \omega\right) + \delta \left( E_{n',\mathbf{k}}-E_{n,\mathbf{k}} - \hbar \omega\right)\right],
\end{equation}
where $e$ is the electron charge, $\Omega$ is the volume of the box,
$N_{\mathbf{k}}$ is the number of points used to describe the electronic bands, $m$ is the mass of the electron, $\hat{\mathbf{M}}$ 
is the squared matrix element that connects the initial and final wave functions of the electrons via the radiation perturbative field, which belong to different bands labelled $n$ and $n'$ but have the same crystal momentum $\mathbf{k}$, while $E_{n,\mathbf{k}}$ is the energy value of band $n$ at the point $\mathbf{k}$ of reciprocal space. Eq. \ref{epsilon} is obtained under the assumption that the transmitted momentum can be neglected (long wavelength limit), but can be extended to finite momentum transfer. The $\delta$ functions are slightly smeared to avoid unphysical excitations with infinite lifetimes. When comparing the spectra of the three boxes with different edges, we found that convergence is achieved with the box with $L=14.26$ \AA, which corresponds to 192 atoms (see Figure S.4 in the SI, where we compare the results for $L=10.28$, $L=14.26$ and $L=21.42$ \AA).\\
\indent 
The calculation of $\epsilon^{(2)}_{\alpha,\beta}$ was performed using DFT as implemented in the Quantum Espresso code suite \cite{qe_2017}.
The ground state electron density was determined with a plane wave cutoff equal to $80$ Rydberg with the Perdew-Burke-Ernzerhof (PBE) functional. Using the same functional, we also investigated the effect of a finite constant electric field on the electronic properties of the pristine matrices and found negligible effects for all amorphous silica configurations.
Once PBE self-consistency is achieved, we have improved the estimated electronic bandgap of pristine amorphous silica by using the HSE06 (Heyd–Scuseria–Ernzerhof) hybrid exchange–correlation functional \cite{hse} (see Figure S.5 in SI), which indeed yields a value of about 8 eV, which is much closer to the experimental values in the range of 8 to 9 eV \cite{exp_sio2,exp_sio2_2,nekrashevich2014electronic}.\\
\indent 
Recall that both sol-gel synthesis and specimen preparation for APT analysis usually result in defects in the pristine amorphous silica matrix, such as vacancies, dislocations and interstitials. In particular, the use of FIB for lift-out can lead to substitution or displacement of atomic constituents by the implantation of Ga$^+$ ions, which can lose all their kinetic energy through elastic and inelastic interactions in the sample and be adsorbed. In the case of gallium, depending on the ion concentration, this can lead to significant changes in the structural and optical properties of the sample. To determine the effects of such defects on the optical properties of the sample, we performed the calculation of the imaginary part of the dielectric function in the long wavelength limit (see equation \ref{epsilon}) for a number of typical defect configurations by including interstitial hydrogen, carbon and sodium atoms, oxygen vacancies \cite{sio2_theor}, interstitial and substitutional gallium ions at different concentrations.
We emphasise that only the positions of the atoms of the defective silica structures were optimised, while the periodic cells were constrained by density; therefore the cubic box has fixed lattice vectors. To optimise the atom positions, we use a BFGS quasi-Newton algorithm that enforces convergence to 10$^{-5}$ Hartree/Bohr and 10$^{-6}$ Hartree for the forces and energy, respectively.
The optical properties for each defected configuration were evaluated using DFT with the HSE06 exchange-correlation functional~\cite{hse06}, which ensures a more accurate evaluation of the resulting absorption properties.   

\section{Results}

\subsection{Experimental results}

APT analyses for chemical characterisation of silica samples with the green laser (532 nm) were often unsuccessful because the samples broke early, generally before 100,000 ions were collected. This is due to the low absorption of silica at this wavelength (see Figure S.2 of the SI), which leads to excessive heating of the sample by the laser. This means that the electrostatic field required for field evaporation cannot be sufficiently reduced. As a result, the applied DC voltage must be increased to achieve the required higher field strengths, leading to premature breakage of the sample due to the induced mechanical stress.  \\
\indent
In contrast, the use of UV or deep-UV light drastically improves the yield of successful runs as a lower onset voltage is required to achieve an acceptable evaporation rate. This allows more data to be collected before the voltage required to maintain evaporation is high enough to cause sample breakage.
The APT samples analysed generally yielded several million ions, which corresponds to a large reconstructed volume.

\begin{figure}[htb!]
   \centering
\includegraphics[width=1\textwidth]{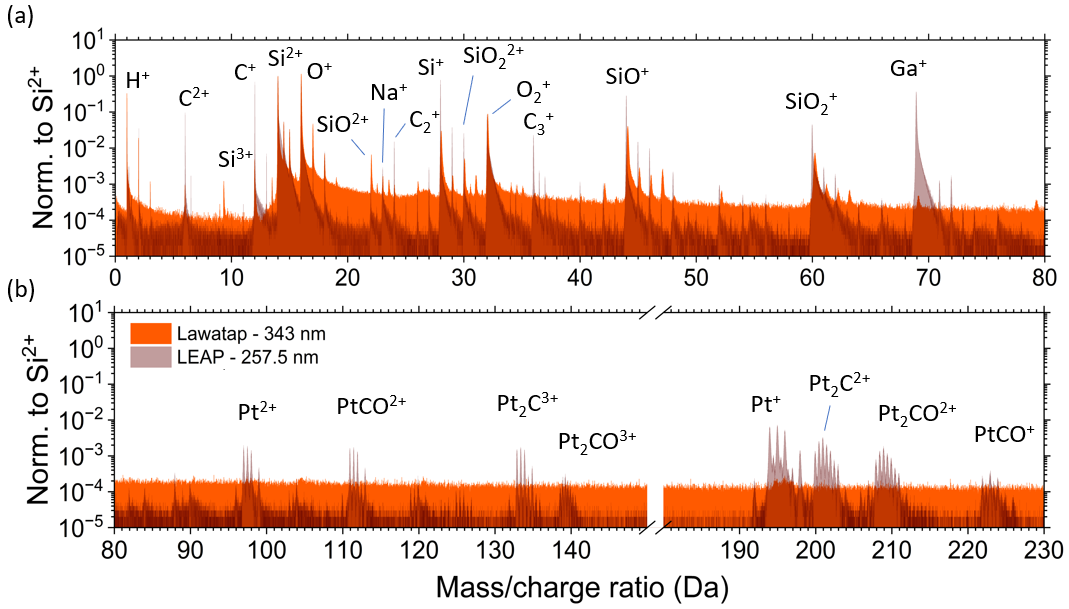}\caption{\label{fig:mass_spectra} Mass spectra of silica samples, measured for the ranges (a) [0-80] and (b) [80-230] of the $m/z$ (Da) values with: Deep-UV (LEAP, dusty pink spectrum) and UV (LaWATAP, orange spectrum); 17 million ions were collected at a laser energy of 250 pJ, corresponding to an energy density of 80 J/m$^2$ for deep-UV, and at 50 nJ, corresponding to an energy density of 70 J/m$^2$ for UV. The detection rate was set to 0.5\% ions/pulse at a base temperature of $T=50$ K for deep-UV and to 0.25\% ions/pulse at a base temperature of $T=50$ K for UV La-APT.}
\end{figure}

The mass spectra obtained with deep-UV light (LEAP, 257.5 nm, dusty pink spectrum) and UV light (LaWATAP, 343 nm, orange spectrum) are shown for the ranges [0-80] and [80-230] of the $m/z$ (Da) values in Figure \ref{fig:mass_spectra}a,b respectively. These spectra are similar: the two main peaks can be safely assigned to Si$^{2+}$ at 14 Da and O$^+$ at 16 Da. The 16 Da peak is identified as O$^+$ and not as O$_2^{2+}$ according to the study of Ref. \cite{bachhav2013investigation}. Other silica-derived peaks in the mass spectrum are attributed to Si$^+$ (28 Da), O$_2^+$ (32 Da), SiO$^+$ (44 Da), SiO$_2^+$ (60 Da) and smaller peaks are identified as SiO$^{2+}$ (22 Da), SiO$_2^{2+}$ (30 Da) and Na$^+$ (23 Da) from the sodium silicate solution. Ga, Pt and various carbonaceous ions, mainly C$^+$ at 12 Da, were detected in all samples originating from the FIB-SEM sample preparation. It can be seen that the detected molecular species vary when the laser energy is varied. In particular, the Si$_2$O$_2^+$ (88 Da) and Si$_2$O$^+$ (72 Da) peaks disappear at low laser energy and higher static field, which is due to the dissociation of molecular ions at higher electric field.\\
\indent The signal-to-noise ratio is better when a deep-UV light is used, with more Pt, Ga and C peaks being detected (see Figure \ref{fig:mass_spectra}b). We emphasise that the two analyses were not carried out under the same field conditions. To obtain an indication of the mean value of the electric field during the analyses, we calculated the charge state ratio (CSR) of silicon as the ratio of Si$^{2+}$ ions to the total number of Si ions. According to Kingham's theory, the CSR is an indicator of the electrostatic field in the entire analysis \cite{kingham1982post}. It should be noted that the Kingham curves for silicon should be treated with caution for a material with properties that differ drastically from pure silicon and the values for the field should be regarded as estimates.
In the deep-UV analysis, the mean value of the CSR was 0.6, which corresponds to a mean field of 19.7 V/nm, and in the UV analysis 0.95, which corresponds to a field of 21.5 V/nm. The evolution of the CSR (and thus the field) during the analyses will be discussed later.
We also note that the detection rates for the analyses in the deep-UV (0.5\%) and UV (0.25\%) range are comparable (recall that in APT analysis, the detection rate can vary by several orders of magnitude, from 0.001\% to 100\%. In addition, the detection rate varies exponentially with the analysis parameters such as the applied voltage and laser energy. Therefore, when setting an automatic detection rate, it is difficult to control its fluctuations with an accuracy of a factor of 2). When transitioning from deep-UV to UV light, the sample temperature must decrease to maintain a similar detection rate, while the electric field is increased from 19.7 to 21.5 V/nm.
This is a first indication that the two illumination conditions heat the sample differently, with the heating being stronger when using the deep-UV light. Furthermore, we point out that the energy densities of deep-UV and UV light are similar (80 and 70 J/m$^2$ respectively). From this we can conclude that the silica samples have a higher absorption coefficient for deep-UV light than for UV light.
Therefore, when using deep-UV light, analyses can be performed at a static field that is lower than when using UV light, reducing the risk of sample breakage due to the mechanical stress caused by the static field.\\
\indent We also notice that working with a lower field also reduces the background noise, as shown in Figure \ref{fig:mass_spectra}a,b for the deep-UV analysis, since the evaporation rate between laser pulses depends on the field when the same value for the base temperature (50 K) is set for both analyses. At these fixed temperature conditions, a reduction in the electric field essentially reduces the emission \cite{lefebvre2016atom}.

\begin{figure}[htb!]
   \centering
\includegraphics[width=1\textwidth]{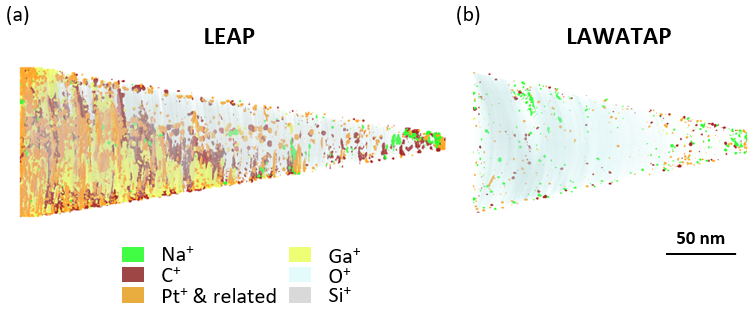}\caption{\label{fig:reconstruction} 3D atomic distribution showing the silica-derived ionic species Si, O (grey and cyan dots, respectively) and C, Na, Ga and Pt (dark brown, green, yellow and orange iso-surfaces with composition thresholds of 18\%, 12\%, 45\% and 25\%, respectively), derived from the FIB-SEM sample preparation, obtained by (a) deep-UV (LEAP) La-APT analysis, reconstructed with a starting radius of 15 nm and a cone angle of 6.5° and (b) UV (LaWATAP) La-APT analysis, reconstructed with a starting radius of 15 nm and a cone angle of 15°.}
\end{figure}

In the reconstructed data shown in Figure \ref{fig:reconstruction}a,b for deep-UV and UV light, respectively, clusters of carbon and platinum were consistently observed in the reconstructed samples. While this is to be expected during sample preparation, where a platinum-carbon mixture is used to protect the sample and attach it to the flat top posts with a GIS, it is usually observed at the beginning or end of the analysis. In this case, however, it is observed throughout the analysis, suggesting that the platinum mixture can fill the micropores in this type of material during sample preparation. It is hypothesised that the precursor gas may diffuse into the pores of the material when the lift-out is welded to the flat top posts. Sodium is also detected throughout the analysis, which is to be expected in the synthesis of silica. The observed distribution of sodium shows a slightly higher atomic concentration at the beginning of the analysis. However, we note that a similar redistribution of another alkali metal, namely lithium, is also observed in films of similar materials due to the electrostatic field \cite{greiwe2014atom} and such effects must be taken into account here, as sodium ions could also migrate through the porous material.
In addition, the diffusion of Pt, Ga and C in the silica volume is more pronounced in deep-UV analysis, probably due to the stronger heating by the laser, which favours diffusion (or, if the silica area is small enough, the laser spot can directly illuminate and heat the Pt weld during APT analysis).\\
\begin{figure}[htb!]
   \centering
\includegraphics[width=0.9\textwidth]{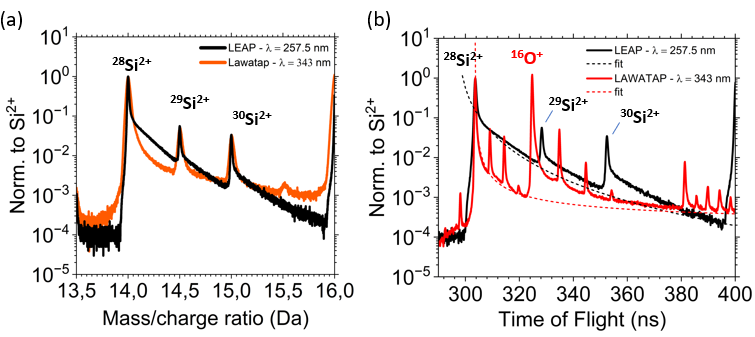}\caption{\label{fig:ToF} (a) Mass and (b) time-of-flight spectra of Si$^{2+}$ ions from the APT analyses of silica samples using deep-UV (black line) and UV (orange line) light. The data sets are taken from Figure \ref{fig:mass_spectra}. The lines in (b) were obtained by fitting with the cooling equation (2), and the values of the cooling time are 7.35 ns and 0.5 ns for the deep-UV and UV analysis, respectively.}
\end{figure}
\indent
To compare the mass resolving power recorded in the UV and deep-UV APT analyses, we show in Figure \ref{fig:ToF}a a zoom of the mass spectra from Figure \ref{fig:mass_spectra}a in the Si$^{2+}$ peak region. The mass resolving power is defined as the ratio between the peak maximum and its width, where the width can be calculated at 10\% and 1\% of the maximum.
The mass resolving power at 10\% is 240 for the deep-UV analysis and 160 for the UV analysis. However, at 1\% the situation is reversed, with a better resolution for the UV APT. Indeed, the tails on the right side of the peaks in the UV spectrum are shorter. The peak tails have been shown to be related to the heating and cooling processes of the sample after laser illumination, causing a delay in the emission of the ions \cite{vella2011field, vella2013interaction}. To analyse this process, it is better to look at the time-of-flight spectra in Figure \ref{fig:ToF}b. Previous studies on La-APT have shown that the cooling time is a function of the heated area, the diffusivity of the sample and its geometry, in particular the cone angle \cite {vurpillot2009thermal,vella2011field}.
The size of the heated area with deep-UV is less than 1 $\mu$m and almost the same size was measured and calculated for UV, taking into account the diffraction of the light at the apex of the sample \cite{vella2013interaction}.
The different cooling times between the two analyses are therefore essentially due to the different geometries of the two samples. From the SEM images, we calculated an average cone angle of 6.5° and 15° for the deep-UV and UV samples, respectively. A change in the cone angle by almost a factor of 3 leads to a change in the cooling time by more than a factor of 10 due to heat propagation \cite{vurpillot2009thermal}.
The tails of the spectra in Figure \ref{fig:ToF}b were fitted using the time-dependent evaporation rate equation \cite{vella2013interaction}:

\begin{equation}
    \Phi(t)=A\cdot \exp \Biggl(-\dfrac{Q}{k_\mathrm{B} \Biggl(T_0+\dfrac{\Delta T_{\mathrm{max}}}{\sqrt{1+\frac{t-t_0}{\tau}}}\Biggr)}\Biggr),
    \label{eq:cooling}
\end{equation}

where $k_\mathrm{B}$ is the Boltzmann constant, $A=0.1\cdot{\exp\Biggl({\frac{Q}{k_\mathrm{B}(T_0+\Delta T_{\mathrm{max}})}\Biggr)}}$ is a normalisation factor, $T_0$ is the base temperature of the tip before laser illumination, which is fixed at 50 K, $\Delta T_{\mathrm{max}}$ is the maximum temperature increase after laser illumination. The tip cooling process starts at $t=t_0$, defined as the point where the Si$^{2+}$ peak intensity falls to 10\% of its maximum, $\tau$ is the cooling time and $Q \approx Q_0(1-\frac{E}{E_\mathrm{evap}})$ is the height of the energy barrier, which depends linearly on the applied electric field $E$ as it approaches the evaporation field $E_\mathrm{evap}$. The values for the cooling time $\tau$ obtained by fitting the tail of the $^{28}$Si$^{2+}$ peak in the time-of-flight spectrum in Figure \ref{fig:ToF}b are 6 ns and 0.5 ns for the deep-UV and UV analyses \cite{vella2011field}. Further details on the fits can be found in the SI (see Figure S.6).\\
\begin{figure}[htb!]
   \centering
\includegraphics[width=1\textwidth]{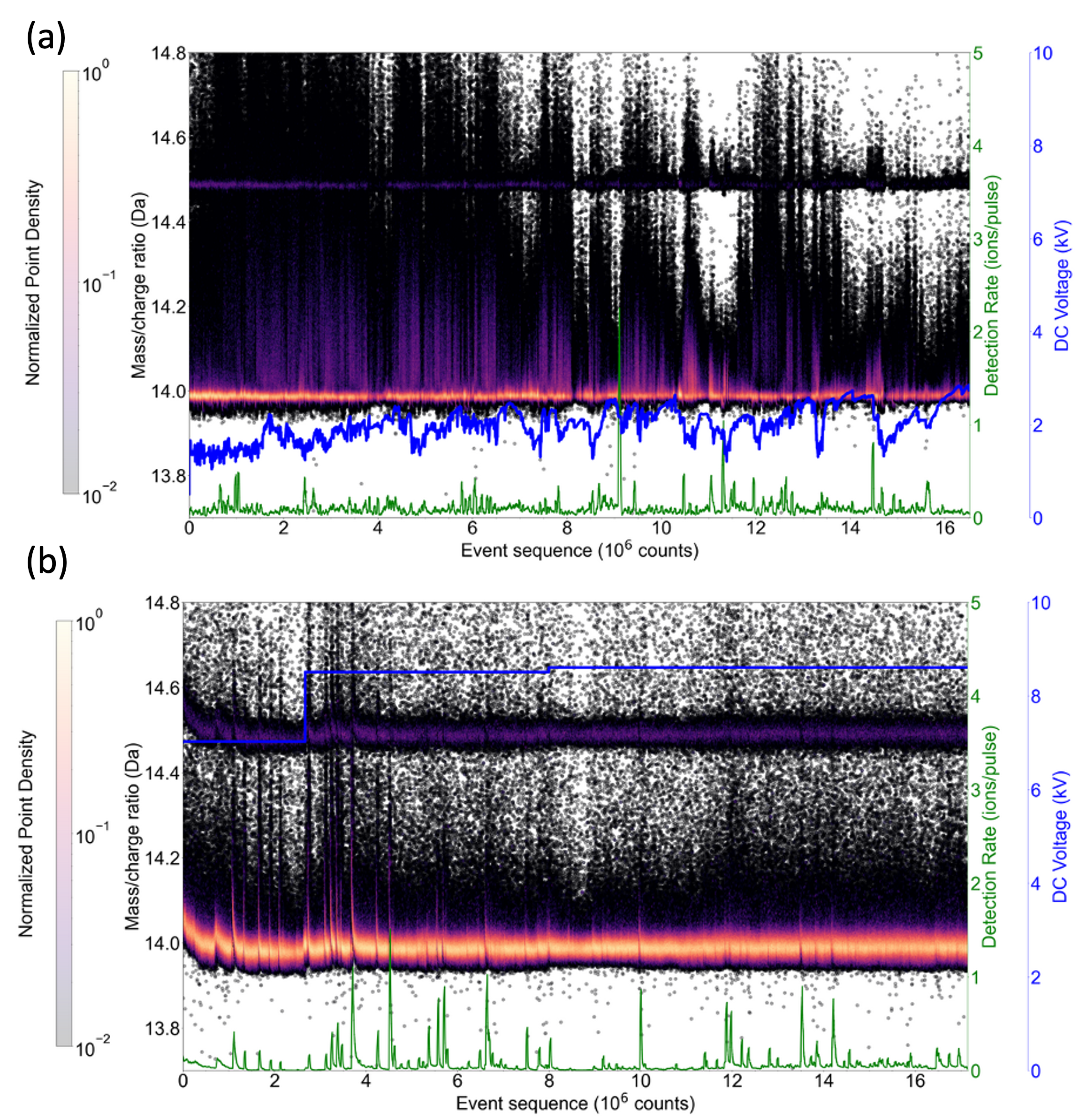}\caption{\label{fig:figure3} History of the variation of the detection rate (green line), the applied voltage (blue line) and the mass spectrum of Si$^{2+}$ ions (dot density plot) as a function of the evaporation sequence for APT silica samples using (a) deep-UV and (b) UV light. The data sets are those shown in Figure \ref{fig:mass_spectra}. The dot density plots of the mass spectrum history are shown on a logarithmic colour scale.}
\end{figure}
\indent Spikes in the detection rate were repeatedly observed in the mass spectra of both deep-UV and UV lasers, as shown in Figure \ref{fig:figure3}a,b. These spikes can be attributed to micro-fractures in the sample, where a small piece breaks off due to the mechanical stress caused by the electric field, resulting in a burst of ion impacts to be detected. This phenomenon can also be partially explained by the presence of mesopores in the sample. If a pore is present, the field to which the surface atoms surrounding the pore are exposed is higher because the effective radius of the sample is smaller. This can lead to a rapid increase in the evaporation rate of the ions. At the same time, the mass-to-charge ratio of Si$^{2+}$ shifts to higher values, as shown in the mass spectrum history plot in Figure \ref{fig:figure3}a,b. This behaviour is more pronounced in UV analysis, which is probably due to the voltage regulation used for deep-UV, which attenuates the fluctuations in the detection rate. 
The shift in the mass spectrum correlated with the evaporation rate has already been reported for materials with a large band gap and low electrical conductivity such as MgO, SiO$_2$ and diamond \cite{arnoldi2014energy,arnoldi2018thermal,caplins2020algorithm,arnoldi2015role}. This is due to the high resistivity of silica, which leads to a voltage drop across the sample that increases with increasing emission current. The voltage drop causes the ions to be emitted at a lower potential so that the mass-to-charge ratio of Si$^{2+}$ in Figure \ref{fig:figure3}a,b shifts to higher values.\\
\indent The resistive behaviour of the silica samples under UV and deep-UV illumination clearly shows that the amount of charge carriers generated by the absorption of light is not sufficient to improve the electrical conductivity of the silica.

\begin{figure}[htb!]
   \centering
\includegraphics[width=1\textwidth]{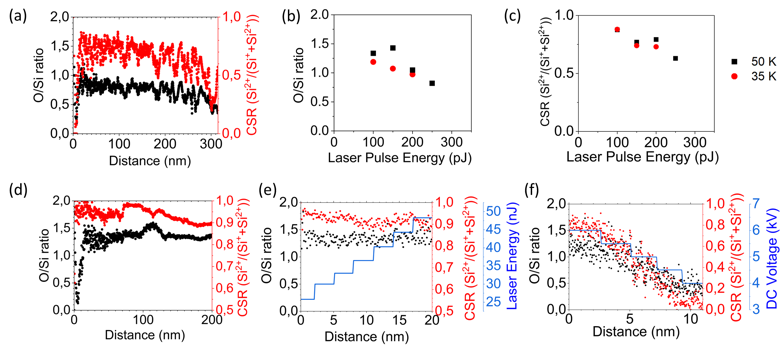}\caption{\label{fig:composition} O/Si atomic ratio (black line) and charge state ratio (CSR, red line) as a function of depth of analysis in silica samples analysed with deep-UV (a) and UV (d) laser illumination.
The evolution of the O/Si ratio (b) and the CSR (c) as a function of laser energy under deep-UV illumination at base temperatures of 30 K (red dots) and 50 K (black dots).
The evolution of the O/Si ratio (black dots, panels (e) and (f)) and the CSR (red dots, panels (e) and (f)) as a function of the depth of analysis in silica samples analysed with a UV laser, where the laser energy (blue line in panel (e)) or the DC voltage (blue line in panel (f)) was varied stepwise.}
\end{figure}

Figures \ref{fig:composition}a,d show the chemical composition obtained from the APT analyses of silica with deep-UV and UV laser, reported in Figure \ref{fig:mass_spectra}. The O/Si ratio was calculated as a function of the depth (or distance) of analysis. This ratio varies between 1.2 and 0.5 when analysed with the deep-UV laser (black dots in Figure \ref{fig:composition}a), but remains almost constant when analysed with the UV laser (black dots in Figure \ref{fig:composition}d).\\
\indent The chemical composition of the oxides in the APT analysis has been shown to depend on the values of the electric field \cite{mancini2014composition}. To test this dependence, we calculated the charge state ratio (CSR) of silica (see red dots in Figure \ref{fig:composition}a,d). Recall that the changes in the CSR during the APT analysis give an indication of the relative changes in the electric field. When compared with the measured composition, it was found that the field in silica was lower in the deep-UV analysis and decreased when Ga, Pt and C were present. In addition, the change in the O/Si ratio exactly follows that of the field, which indicates that silica exhibits the same behaviour as other materials with a large band gap. In the UV analysis, the field remains almost constant, as does the composition with an O/Si ratio of about 1.4 (black dots in Figure \ref{fig:composition}d). \\
\indent We have also investigated the dependence of the composition on the energy of the laser pulse. When using LEAP APT (deep-UV light), the analyses are performed at a constant detection rate, so that an increase in laser energy is always accompanied by a decrease in the applied voltage and thus the field. Figure \ref{fig:composition}b,c show the dependence of the O/Si ratio and the CSR on the laser energy. These two quantities decrease with increasing laser energy with almost the same gradient. This means that the dependence of the composition on the laser energy is negligible. For the UV analyses, it was possible to change the laser energy and the voltage separately. The results are shown in Figure \ref{fig:composition}e,f. When the laser energy was increased from 12 to 48 nJ, the O/Si ratio increased by 10\%, but when the voltage (i.e. the CSR) was increased from 4 to 6 kV (CSR from 0.08 to 0.8), the O/Si ratio increased by a factor of 3 from 0.4 to 1.2, clearly showing the strong dependence of the composition on the field values.
For both deep-UV and UV laser illumination, the measured composition deviates from the expected nominal composition (O/Si ratio of 2), but the measured ratio approaches this value at higher electric fields, which corresponds to a higher CSR of Si.

\subsection{Theoretical results}

\begin{figure}[hbt!]
   \centering
\includegraphics[width=1.0\textwidth]{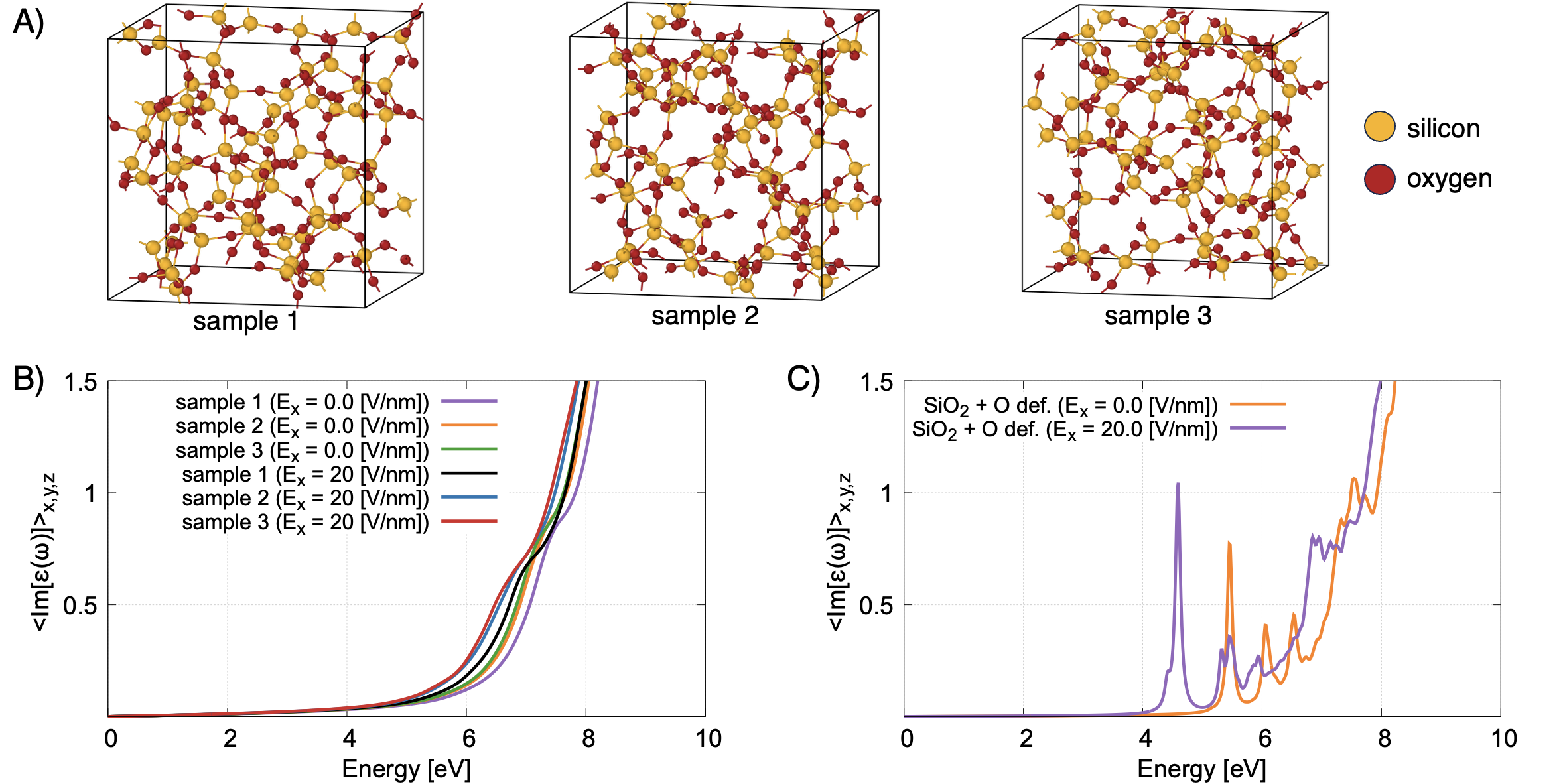}
\caption{\label{fig:figureSI5a-SI5b}(a) SiO$_2$ simulation supercells with $N=192$ atoms, all corresponding to an edge of the box $L=1.42$ nm. There are no defects in these structures. The oxygen atoms are shown in red, the silicon atoms in yellow. (b) Influence of the static electric field: average of the imaginary part of the dielectric tensor along the orthogonal Cartesian directions for samples 1 (left in the panel (a) above), 2 (centre), 3 (right). Pristine amorphous silica matrix without electric field (violet, orange and green lines) and with a constant electric field of 20 V/nm along the $x$ direction (black, blue and red lines). (c) Amorphous silica matrix with a non-bridging oxygen defect without electric field (orange line) and with a constant electric field of 20 V/nm along the $x$ direction (violet line). These calculations were performed with a PBE functional that notoriously underestimates the fundamental gap.}
\end{figure}

\begin{figure}[htb!]
   \centering
\includegraphics[width=1\textwidth]{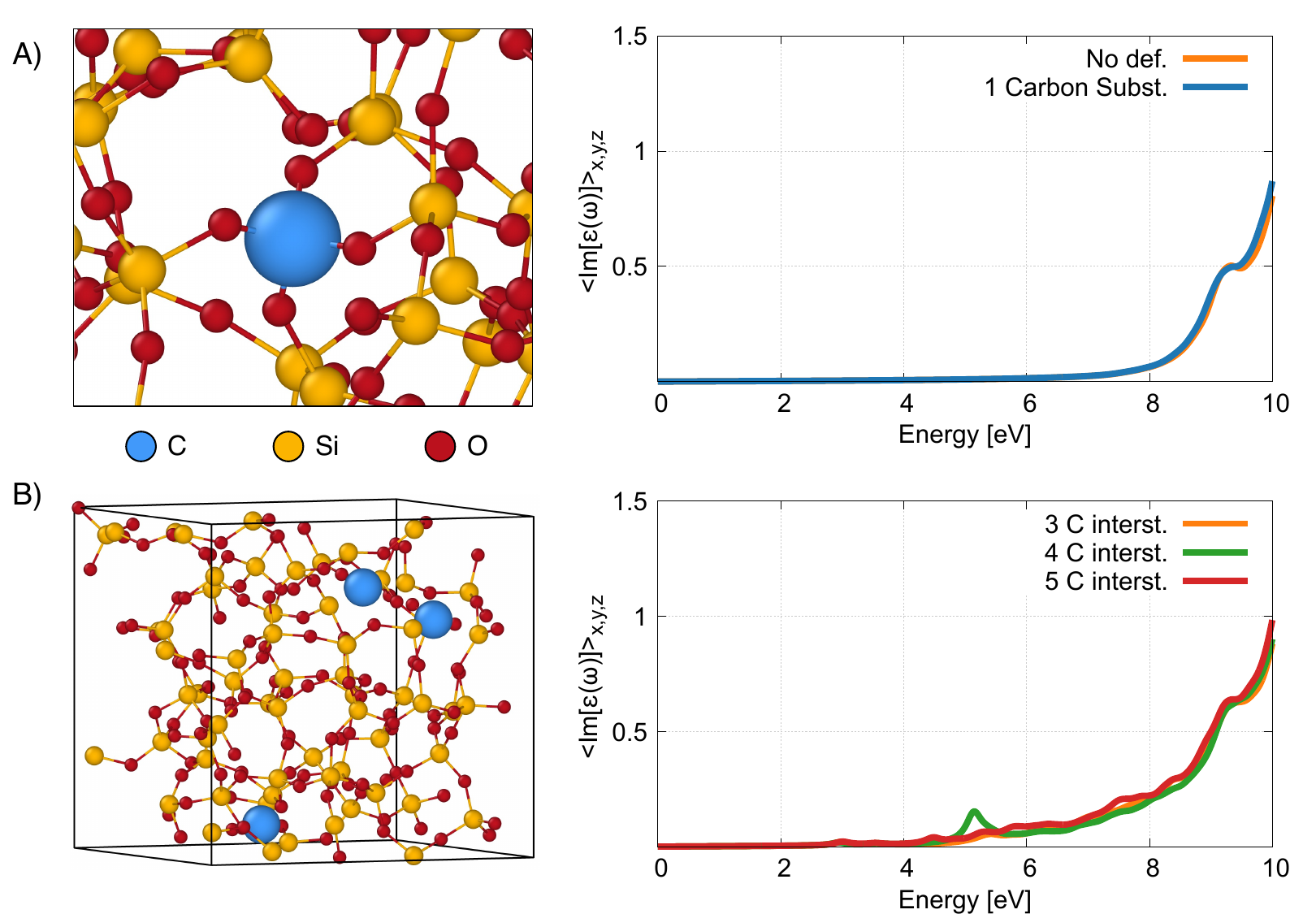}\caption{\label{panel_abs_spectr_C} (a) Absorption spectra (right panel) of the silica sample shown in the left panel for one substitutional carbon atom (in blue colour) replacing silicon (yellow colour); (b) absorption spectra (right panel) of the silica samples with three (as shown in the left panel) to five interstitial carbon atoms (blue colour in the left panel of the figure).}
\end{figure}

The atomic distribution in Figure \ref{fig:reconstruction} shows that a number of impurities can be detected in both UV and deep-UV analyses, which can dramatically change the optical properties of silica \cite{PhysRevB.57.818,PACCHIONI1999175}.
In this context, we note that, in principle, three effects can interact to reduce the band gap: firstly, the presence of a strong electrostatic field. Secondly, phonon excitation and/or lattice distortion due to the heating of the sample during irradiation. The bandgap energy of semiconductors or insulators such as silica, decreases with increasing temperature as the amplitude of atomic vibrations increases, leading to larger interatomic distances. In this work we only perform static single-point calculations of the absorption spectra on optimised geometries with defects, i.e. the nuclei are clamped and the strong static and time-dependent fields acting on the samples, which in principle have an influence on the change of the equilibrium bond length due to the ionic charges, only affect the electronic degrees of freedom. Third, the presence of defects, which typically lead to the generation of intragap states.
In particular, we analyse two types of defects: Substitutional defects, in which an atom of the silica matrix, be it Si or O, is replaced by another type of atom, such as carbon or gallium, but occupies the same position; and interstitial defects, i.e. crystal defects in which an atom of (the same or) a different type occupies an interstitial site in the crystal structure (e.g. interstitial gallium defects are gallium atoms that occupy a site in the crystal structure of silica where normally no atom is present).\\ \indent To obtain a convincing interpretation of the role of defects in the absorption properties of amorphous silica in the APT setup and to explain the different emission rates for the frequency ranges investigated in this work, we have determined the ab initio optical absorption spectra for three different DFT-optimised models of pure amorphous silica (see Figure \ref{fig:figureSI5a-SI5b}a), all characterised by a cubic box of 1.426 nm size with 192 atoms at a density of 66.2 atoms/nm$^3$ and representing relative minima in the potential energy landscape. Figures S.7 and S.8 of the SI also show the real and imaginary parts of $\epsilon(\omega)$ and the refractive index, which lies between 1.5 and 1.7 in the energy range of interest, for three defect-free amorphous structures to which we add a number of representative defects each time.\\ 
\indent Atomic evaporation in APT analysis takes place under static and external laser fields. The effect of strong static fields on the optical absorption properties of silica matrices was therefore initially investigated for a constant electric field of 20 V/nm applied in the horizontal direction. Note that the value of the field is close to the experimental values calculated from the CSR for the two analyses. The optical properties for each configuration were evaluated using DFT with the PBE exchange-correlation functional. We note that the 0 of the energy axis coincides with the top of the valence band in all calculated absorption spectra. A consistent red shift of the optical absorption spectrum was observed in all three samples compared to a zero electrostatic field (see Figure \ref{fig:figureSI5a-SI5b}b), which leads to a slight reduction in the band gap of the amorphous silica. However, this shift is relatively small and did not significantly change the overall absorption properties. These results indicate that the applied static field does not fundamentally disrupt the optical response of amorphous silica, but rather causes subtle systematic perturbations.
However, if a bridging oxygen is removed from the amorphous silica structure, the red shift of the spectrum is $\simeq 1$ eV (violet line in Figure \ref{fig:figureSI5a-SI5b}c), which is more pronounced than in the case without defect (see orange line in Figure \ref{fig:figureSI5a-SI5b}c).
This indicates that the defects play an extremely important role in tuning the light absorption.\\
\indent Motivated by this indication and driven by our experimental results, we investigated a variety of typical defects in the silica matrix, which can also occur during the sol-gel and lift-out process, to interpret the effective laser light absorption in the UV and deep-UV range. In particular, we have focussed on interstitial C, H, O and Na atoms as well as interstitial and substitutional gallium atoms with or without, near or far oxygen vacancies. The optical properties for each configuration were evaluated using DFT with the HSE06 exchange-correlation functional, as it reproduces the experimental bandgap of amorphous silica much more accurately than PBE (see Figure S.5 in the SI).\\
\indent
In Figure \ref{panel_abs_spectr_C}a we show the absorption spectra of such optimized defective silica structures obtained by replacing one silicon atom with a substitutional carbon atom (blue atom in the figure on the left). We have tried different configurations for the carbon atoms and different amorphous silica matrices without finding significant discrepancies. In Figure \ref{panel_abs_spectr_C}b we show the absorption spectra when three to five interstitial carbon atoms (blue colour on the left side in Figure \ref{panel_abs_spectr_C}b) are added to the original pristine silica matrix.

From the analysis of the absorption spectra in Figure \ref{panel_abs_spectr_C}a,b we conclude that substitutional carbon atoms in different positions do not significantly affect the absorption properties of the silica matrix (see right panel in Figure \ref{panel_abs_spectr_C}a). This conclusion could also be drawn from the observation that the valence band edge of SiO$_2$ consists mainly of non-bonding $O2p$ orbitals \cite{nekrashevich2014electronic}. Substitution with carbon does not drastically change the band gap, which is consistent with the calculated absorption spectra.
However, interstitial carbon defects at different concentrations and coordination numbers increase the intragap light absorption of silica between 4 and 8 eV (see right panel in Figure \ref{panel_abs_spectr_C}b), which could explain the stronger absorption observed experimentally in the deep-UV range.\\
\begin{figure}[htb!]
   \centering
\includegraphics[width=1.0\textwidth]{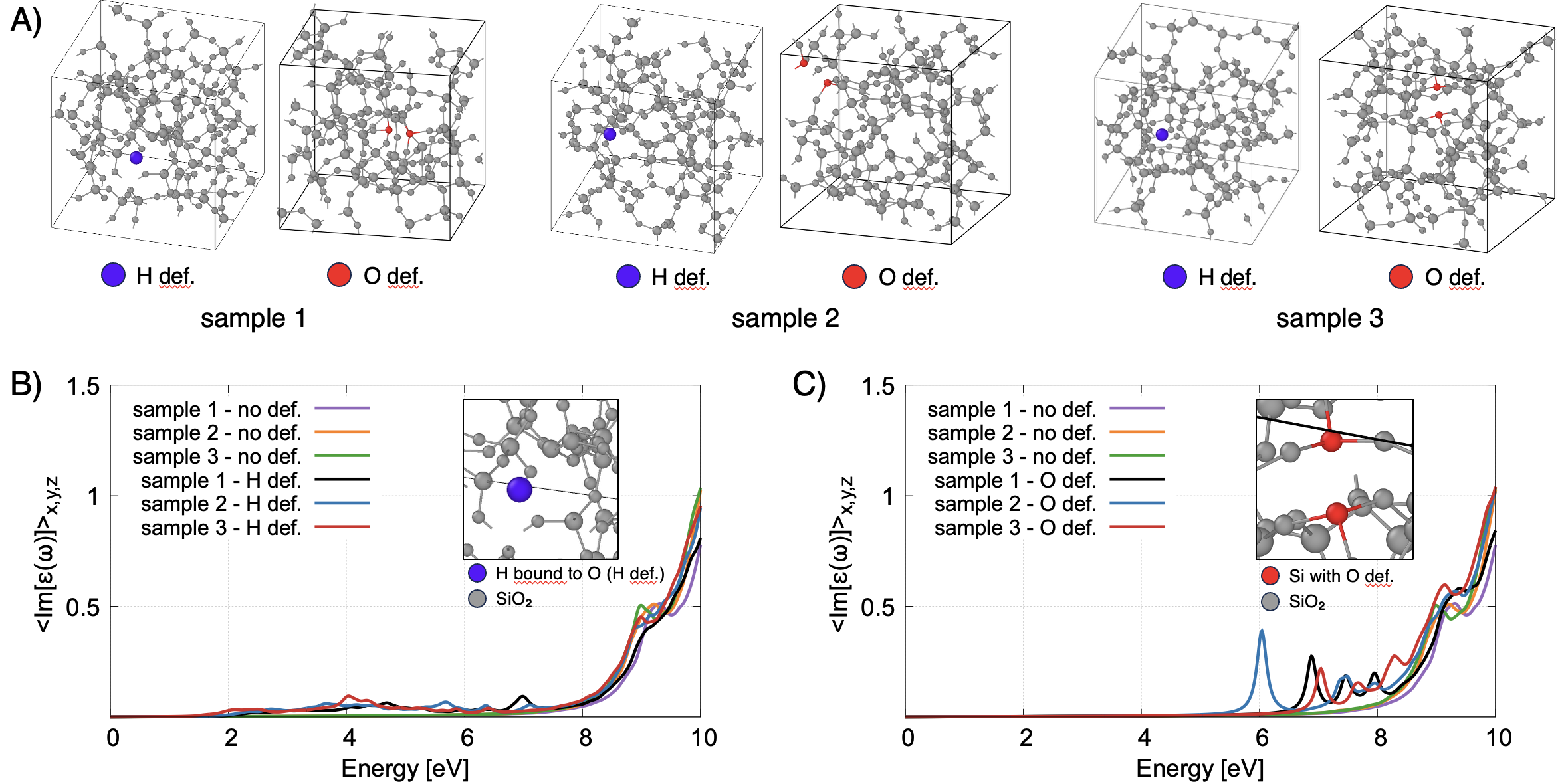}\caption{\label{panel_abs_spectr_H} Absorption spectra of the silica samples shown in panel (a), for (b) interstitial hydrogen (H def., blue colour) bound to oxygen and (c) for an oxygen vacancy (O def., red colour) in different configurations compared to the defect-free amorphous structures of Figure \ref{fig:figureSI5a-SI5b} (no def.).}
\end{figure}
\indent The behaviour of the absorption lineshape also changes considerably when either an interstitial hydrogen atom bound to oxygen or a bridging oxygen vacancy is introduced in different configurations of the three representative silica samples (see Figure \ref{panel_abs_spectr_H}a).
Indeed, in both cases we observe the appearance of several structured intra-gap absorption peaks between 4 and 8 eV upon the insertion of an interstitial hydrogen atom (see Figure \ref{panel_abs_spectr_H}b) and between 6 and 8 eV upon the creation of a bridging-oxygen vacancy (see Figure \ref{panel_abs_spectr_H}c). This goes towards the direction of the experimental results of lowering the band gap and allowing intra-gap absorption.\\ 
\indent We have also analysed the effects of the likely presence of atomic and molecular sodium as well as gallium from the lift-out process in the silica matrix, even in the presence of oxygen vacancies. Previous atomistic computer simulations provided a comprehensive understanding of the structure of sodium silicate glasses and showed that sodium atoms tend to cluster in alkali-rich regions \cite{10.1063/1.459296,10.1063/1.460814}, which we modelled by a sodium molecule within the matrix (see left side of Figure \ref{panel_abs_spectr_Na}a), and to aggregate around non-bridging oxygen atoms, which we modelled by a sodium molecule inside the matrix near oxygen vacancies (see left side of Figure \ref{panel_abs_spectr_Na}b). 
\begin{figure}[htb!]
   \centering
\includegraphics[width=1\textwidth]{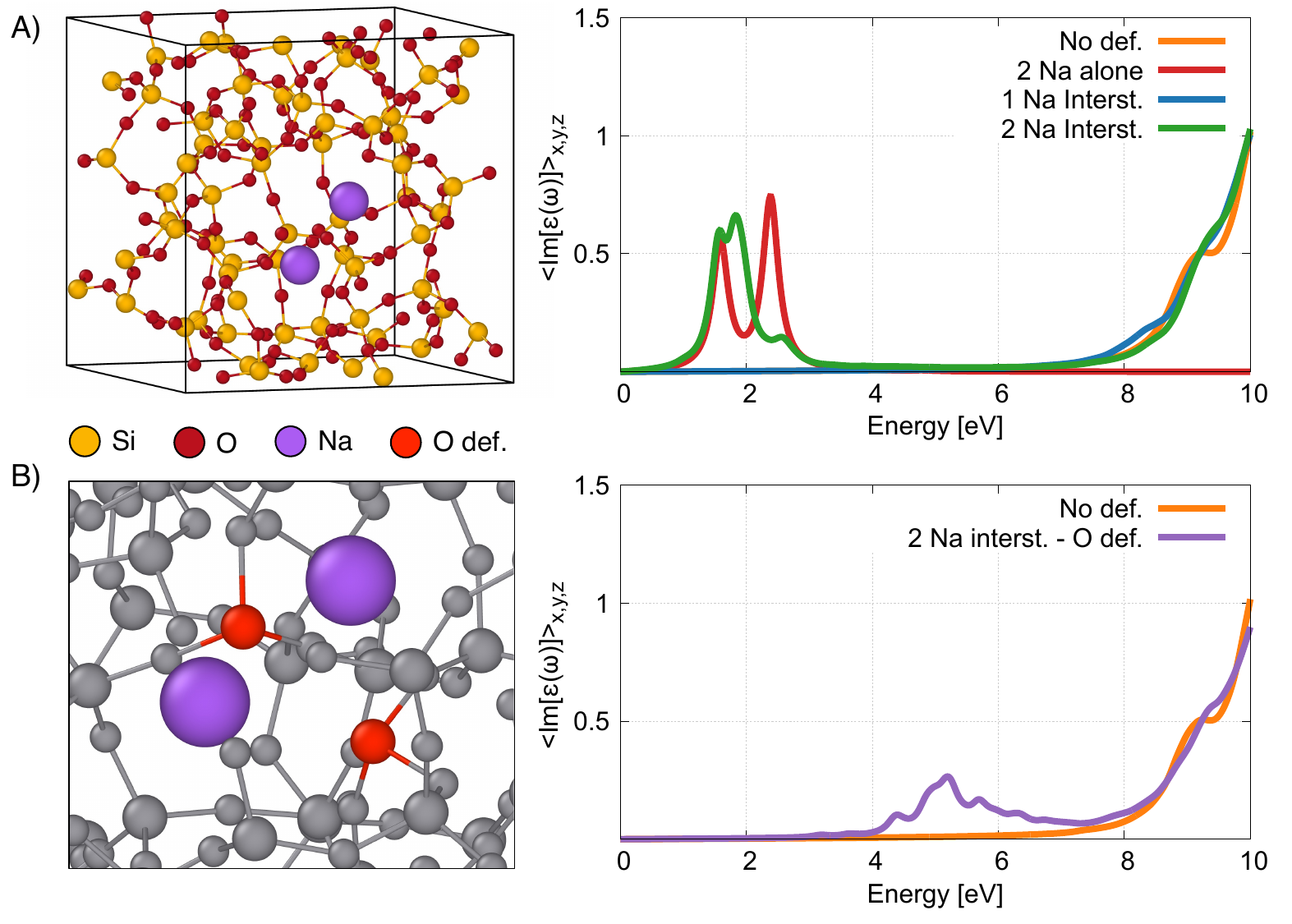}\caption{\label{panel_abs_spectr_Na}(a) Left: optimised configuration for interstitial sodium defects (violet atoms) in the amorphous silica matrix. Right: absorption spectra of the silica samples with and without (orange line) the interstitial sodium atoms (blue and green lines) and of molecular sodium (red line).
(b) Left: optimised configuration for molecular sodium near a bridging oxygen vacancy inside the amorphous silica matrix. Right: Absorption spectrum of the amorphous silica sample in the simultaneous presence of a sodium molecule in the vicinity of a bridging oxygen vacancy (violet line) and in pristine (defect-free) form (orange line).}
\end{figure}
The presence of interstitial atomic or molecular sodium does not significantly change the absorption spectra of the silica matrices (see right side of Figure \ref{panel_abs_spectr_Na}a).
In fact, we could not observe any defect states within the optical gap for interstitial sodium atoms, as the absorption spectra are essentially the mere overlap of the spectra of atomic (blue and green lines in the right panel of Figure \ref{panel_abs_spectr_Na}a) or molecular sodium (red line in the right panel of Figure \ref{panel_abs_spectr_Na}a) and the defect-free silica matrix (yellow line in the right panels of Figure \ref{panel_abs_spectr_Na}a,b).
However, by introducing bridging oxygen vacancies into the matrix (see left side of Figure \ref{panel_abs_spectr_Na}b), the sodium cluster aggregates near these vacancies and we observed an increase in absorption at about 4-6 eV (see right side of Figure \ref{panel_abs_spectr_Na}b), which is in the range of our experimental results with deep-UV light at about 4.8 eV.\\
\indent 
\begin{figure}[htb!]
   \centering
\includegraphics[width=1\textwidth]{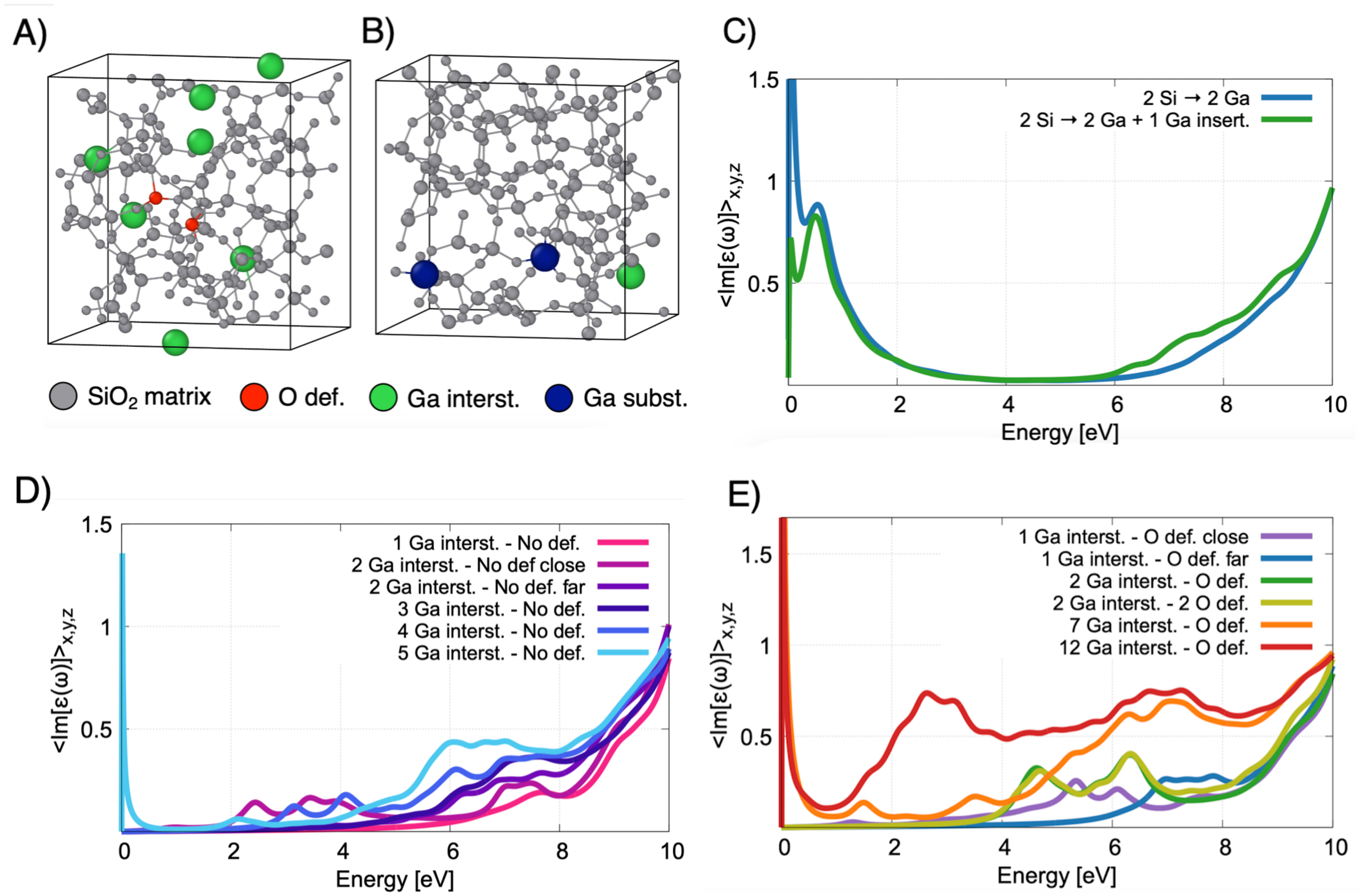}\caption{\label{panel_abs_spectr_Ga} (a) Model of a silica sample with the simultaneous presence of several interstitial gallium atoms (green), one and two oxygen vacancies (red); (b) model of a silica sample with the simultaneous presence of two substitutional (blue) and one interstitial (green) gallium atom.
(c) Absorption spectra of silica samples with two substitutional gallium atoms (blue line) also in the presence of interstitial gallium atoms (green line).
(d) Absorption spectra of different silica samples with different numbers of interstitial gallium atoms without oxygen defects, at least in the vicinity.
(e) Absorption spectra of different silica samples with different numbers of interstitial gallium atoms, which may be close to or far from the oxygen vacancy. 
}
\end{figure}
Finally, we have analysed the change in light absorption properties due to the presence of interstitial (see green atoms in Figure \ref{panel_abs_spectr_Ga}a) and substitutional (see blue atoms in Figure \ref{panel_abs_spectr_Ga}b) gallium atoms, also in the presence of bridging oxygen vacancies. Our simulations show that the substitution of silicon by gallium atoms in the silica matrix leads to the formation of metallic states at the Fermi level (see green and blue lines around 0 eV in Figure \ref{panel_abs_spectr_Ga}c), which makes the material conductive. This behaviour is in stark contrast to experimental observations in which gallium-doped silica retains its insulating character. Consequently, a simple substitution model for the incorporation of gallium appears to be an unlikely representation of the actual atomic configuration in our samples.\\
\indent
Nevertheless, the presence of interstitial gallium defects leads to new intra-gap electronic states (see Figure \ref{panel_abs_spectr_Ga}d), which contribute to increased optical absorption, especially at low photon energies. The simulations show that increasing the number of Ga atoms in the cubic box above 5 (i.e. at a concentration of about 2\%) leads to the formation of metallic states at the Fermi level (see cyan line in Figure \ref{panel_abs_spectr_Ga}d), which makes the material conductive at odds with our experimental observations. For 4 interstitial Ga atoms in the box, our calculations indicate a significant increase in absorption from energies of 3.5 eV (see blue line in Figure \ref{panel_abs_spectr_Ga}d), even in the absence of oxygen defects, which could explain the experimentally observed features in the UV and deep-UV regions. These defect-induced states may play a role in modifying the local electronic structure without fully metallising the material, depending on the distribution and concentration of gallium defects in the host matrix. Finally, in Figure \ref{panel_abs_spectr_Ga}e we show the absorption lineshape when one (O def.) or two (2 O def.) bridging oxygen vacancies are present near (O def. close) or at a distance (O def. far) from interstitial gallium atoms, up to a concentration of 6\% in Ga. We conclude that the presence of interstitial gallium atoms together with bridging oxygen vacancies in different configurations (see Figure \ref{panel_abs_spectr_Ga}e) could also explain the light absorption in the UV and deep-UV range, in agreement with our experimental results. 

\section{Discussion and Conclusions}

The outcome of our theoretical and computational study is that the better performance of UV and deep-UV La-APT compared to green La-APT could be due to the higher absorption of the silica matrices in the shorter wavelength range due to defects of different nature, such as oxygen vacancies also in the presence of interstitial impurity atoms from synthesis processes.\\
\indent
A higher absorption leads to the generation of more free charges and thus to a higher energy transfer into the structure, which in turn leads to a higher temperature of the silica matrix. This means that a high electric field is not required to evaporate the surface atoms, reducing the risk of sample breakage.
When many free charges diffuse in the sample, the resistivity of the silica also decreases, so the potential drop along the tip is also reduced due to the resistivity effect. This makes it possible to increase the effective electric field at the surface and thus evaporate the atoms more easily.
The absorption spectrum of sol-gel silica in Figure S.2 of the SI actually shows a slight increase at wavelengths below 400 nm and a stronger increase at wavelengths below 300 nm. In particular, the absorption between 343 nm (UV light) and 258 nm (deep-UV light) increases by a factor of almost three. This increase in sub-bandgap absorption at 258 nm is a clear indication of the presence of defects in the silica.
DFT calculations performed on silica matrices with interstitial carbon atoms or molecular sodium with oxygen vacancies confirm that their presence increases the sub-bandgap absorption for deep-UV light, but not for UV photons.\\
\indent
The experimental results were obtained with deep-UV and UV La-APT at similar laser energy density, but different field and thus different temperature, indicating a lower absorption of UV light than deep-UV light by almost a factor of 4. However, the DFT spectra show very low absorption at 3.6 eV, even when high concentrations of carbon and sodium defects are taken into account.\\
\indent 
Since the optical properties of silica can be altered during nanotip fabrication by FIB-SEM milling, as already observed for Si and GaN \cite{bogdanowicz2018laser}, we also need to consider the effects of gallium defects at different concentrations in the silica specimen. In addition, the enormous electric field can increase the absorption by reducing the band gap of the silica in a small region near the apex where the field is very high.\\
\indent 
DFT calculations show that the electric field has only a minor influence on the optical properties of the amorphous silica matrix, but that the presence of a high interstitial gallium density actually increases UV absorption.\\
\indent 
The presence of substitutional gallium or a high density of interstitial gallium makes the material metallic, which is very unlikely in our samples. In fact, this result contradicts the experimental observation of the high resistivity of the material, which leads to flux-dependent voltage drops along the tip, as shown in Figure \ref{fig:figure3}.\\
\indent 
In summary, in this work we have analysed sol-gel silica with La-APT under green, UV and deep-UV illumination. We have shown that although the analysis with green laser-assisted APT was not successful, the use of high-energy photons in the UV and deep-UV range significantly improves the success rate and makes it possible to obtain a large-volume, three-dimensional image of the sample. Furthermore, UV and deep-UV light allows us to work with a lower static field, reducing the mechanical stress on the sample and the risk of breakage. When we compare the results of UV and deep-UV light, we also find that the latter heats the sample more (with similar illumination conditions in terms of laser intensity); therefore, the deep-UV analyses can be performed with a lower static field, which improves the signal-to-noise ratio in the mass spectrum. However, working with a lower field and the associated higher heating increases the diffusion of gallium, platinum and carbon ions in the pores of the sol-gel silica matrix.\\
\indent We have also shown that the evaporation of sol-gel silica is irregular, with some bursts of evaporation probably due to the porosity of the material. The abrupt fluctuations in the evaporation rate cause an energy deficit in the emitted ions due to the low electrical conductivity of the sample. Illuminating the sample with UV or deep-UV light increases the density of the free charge carriers, but is not sufficient to completely eliminate the problem of low conductivity. \\
\indent 
As far as the chemical composition is concerned, we report deviations from the nominal composition, but the measured values are closer to the nominal values at a high electric field. Due to the strong local heating caused by the absorption of laser energy by the defects, the three-dimensional resolution and chemical composition are potentially limited.\\
\indent To investigate the optical absorption properties of silica matrices even in the presence of defects, we have applied density functional theory with the HSE06 hybrid exchange-correlation functional for a variety of defects, such as interstitial and substitutional carbon, hydrogen, sodium, and gallium atoms in different concentrations and configurations. Our thorough defect analysis has highlighted four particularly relevant cases: (i) the oxygen vacancy that disrupts the Si–O–Si network by breaking a Si–Si bond; (ii) the interstitial hydrogen defect that passivates and weakens the oxygen bonds; (iii) an oxygen vacancy with a Na$_2$ molecule nearby; (iv) interstitial and substitutional gallium atoms, which could originate from the synthesis process of the needle-shaped silica specimens for the APT analyses, also in the presence of bridging oxygen vacancies. In the latter configuration, where the gallium atoms populate the matrix with a concentration of 2\%, which can be even lower if oxygen vacancies are included in the model, absorption features around 3.5-5 eV clearly emerge, indicating a signature of the defects in the experimental spectra and their role in enabling APT analyses of silica with UV and deep-UV light.\\
\indent We have also investigated the influence of strong static electric fields comparable to those used in APT analysis and found that fields of the order of 20 V/nm do not significantly alter the optical absorption spectrum. This indicates a high degree of robustness of the optical response of silica under such conditions.\\
\indent
Further improvements of the theoretical description — especially to capture excitonic effects - would require the application of post-DFT methods such as the Bethe–Salpeter equation (BSE). However, the computational cost of these advanced techniques is prohibitively high for systems of this size. In addition, further investigations, possibly involving more complex defect geometries, are required to clarify the exact role of defects in altering the electronic and optical properties of silica-based systems.

\begin{acknowledgement}

This action has received funding from the European Union under the Mimosa grant agreement No 10104665. Part of this work was carried out at Chalmers Materials Analysis Laboratory (CMAL).
The authors gratefully acknowledge the use of the HPC facilities at UniTN and FBK.

\end{acknowledgement}

\begin{suppinfo}

\section*{Pores analysis}
\refstepcounter{figure}
\renewcommand{\thefigure}{S.\arabic{figure}}

The isotherm of the silica sample is shown in Figure \ref{fig:figureSI1}. From this it can be deduced that the material contains both micropores and mesopores. The hysteresis loop observed in the multilayer range of the nitrogen sorption measurement is typical of silica gels and porous glasses. This hysteresis is due to a difference in the relative pressure required for adsorption or desorption. This is generally caused by a capillary condensation effect that depends on the pore radius of the material \cite{Sing+1985+603+619}. An interpretation of this shape is to attribute it to the network effects of pores with narrow necks and wider bodies (in the form of ink bottles), where there is blockage of the pores that remain filled until the necks are emptied during desorption \cite{Thommes}.
It was found that the micropores have an average width of 6.8 \AA~and the mesopores have an average size of 40 \AA.

\setcounter{figure}{0}
 \begin{figure}[htb!]
   \centering
\includegraphics[width=0.8\textwidth]{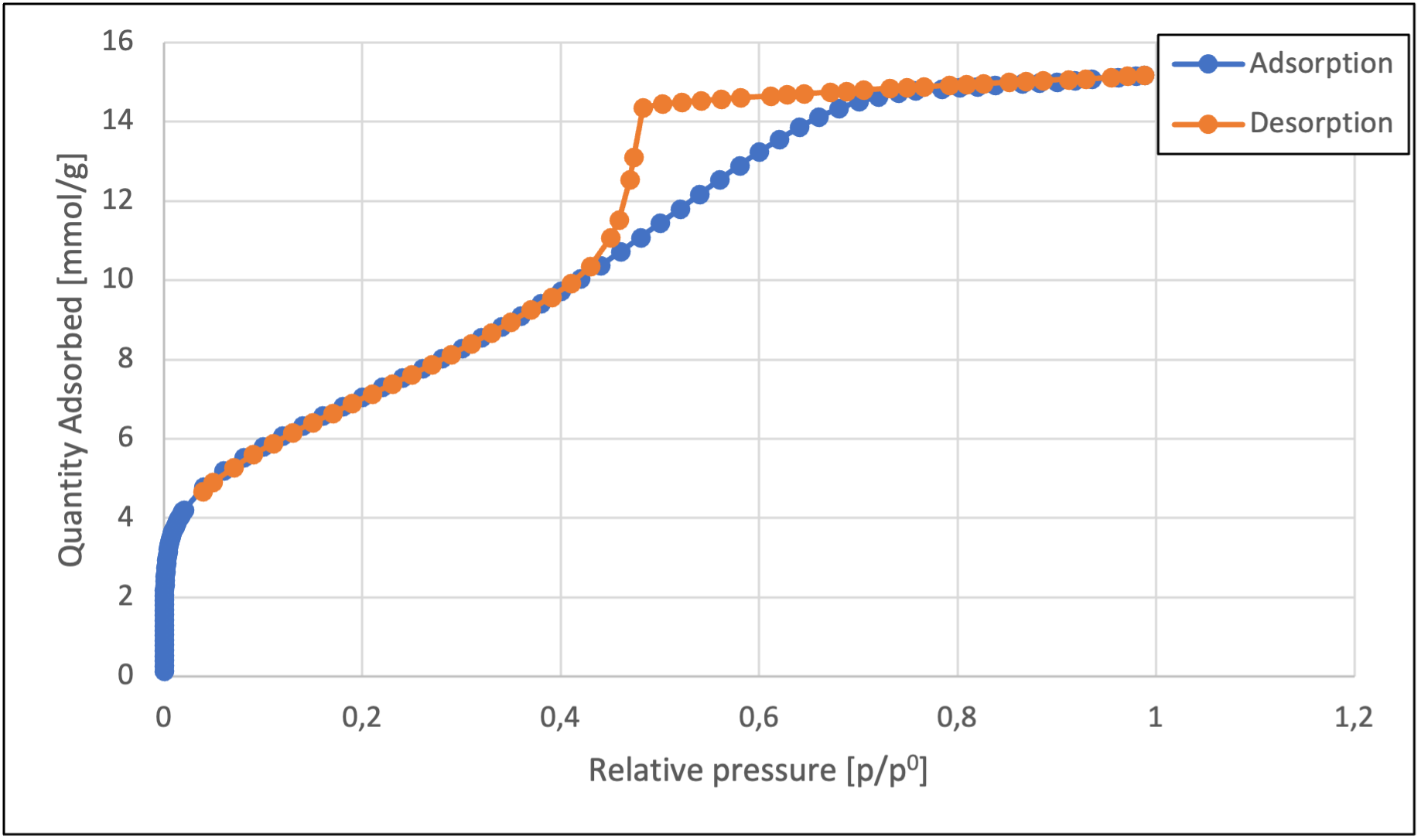}\caption{\label{fig:figureSI1} Complete isotherm measured on a representative silica sample.}
\end{figure}

\section*{UV/Vis spectroscopy of silica}

The absorbance of a thin silica sample in the UV/Vis range was measured with a Hewlett-Packard 8453 UV/Vis spectrometer, which was blanked against air. The absorption spectrum obtained is shown in Figure \ref{fig:figureSI2}. It can be seen from the spectrum that the relative absorbance increases gradually below 400 nm and more rapidly below 300 nm. The constant measured absorbance above 400 nm is in part due to the scattering of light by the sample in addition to the absorption. Note that the sample is not perfectly flat, which contributes to the scattering of light.
\begin{figure}[htb!]
   \centering
\includegraphics[width=0.8\textwidth]{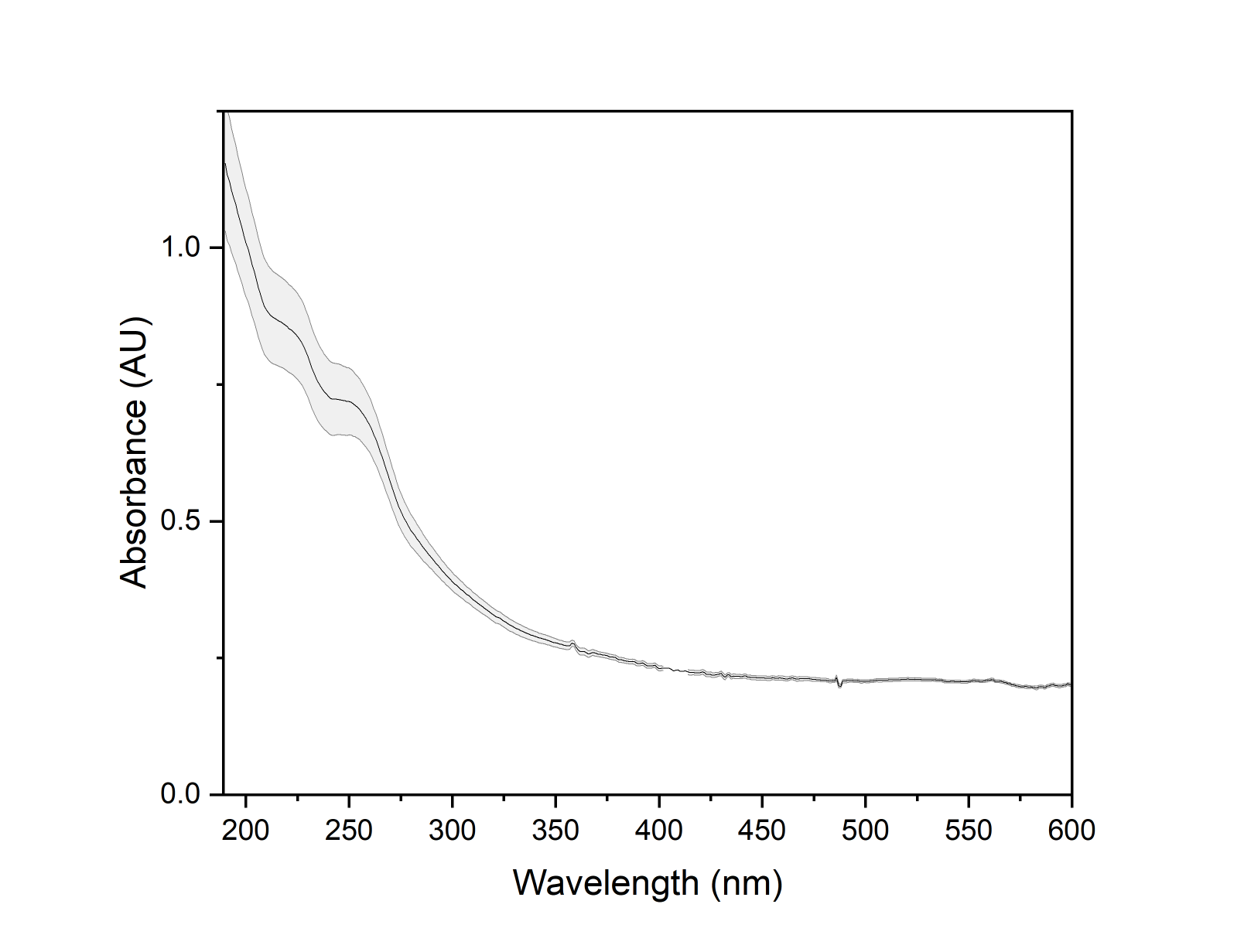}\caption{\label{fig:figureSI2} Absorption spectrum of a 0.6 mm thick silica sample that has been blanked against air. The standard deviation is shown in grey.}
\end{figure}

\section*{Computational details of absorption spectra}
\begin{figure}[hbt!]
   \centering
\includegraphics[trim={0 2.5cm 0 1cm},clip,width=0.9\textwidth]{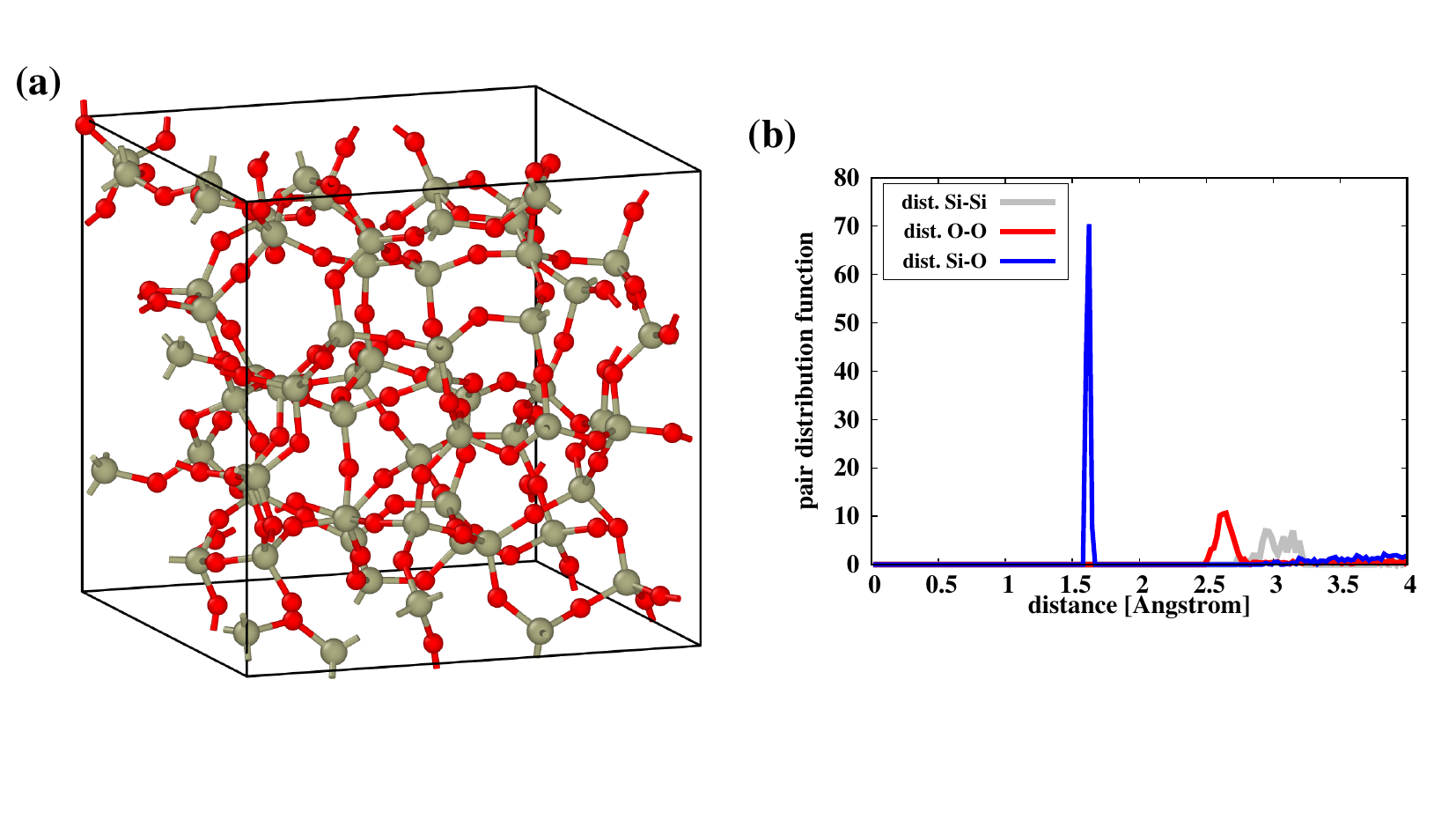}
\caption{\label{fig:figureSI3} Model of the amorphous silica matrix: (a) SiO$_2$ cell with $N=192$ atoms, corresponding to an edge of the box $L=1.42$ nm. There are no defects in this structure. The red atoms are oxygen atoms, while the grey atoms are silicon atoms. (b) Corresponding pair distribution function.}
\end{figure}

In Figure \ref{fig:figureSI3}a we show a model of an amorphous silica matrix that we created using the MC method described in section 2.3 of the main text of the manuscript. In Figure \ref{fig:figureSI3}b we show its pair distribution with the typical peak centred at 1.6 \AA~representing the distance between Si and O atoms in amorphous silica matrices.
\begin{figure}[hbt!]
   \centering
\includegraphics[width=0.8\textwidth]{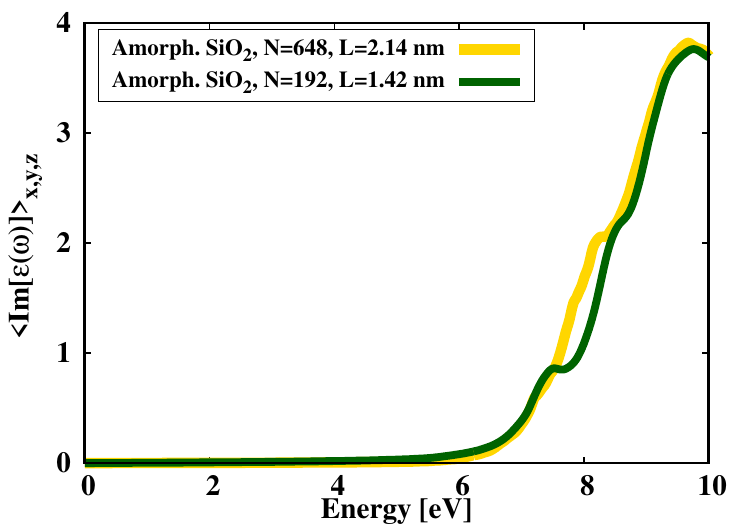}
\caption{\label{fig:figureSI4} Effect of the finite size of the calculation supercell on the imaginary part of the dielectric tensor, averaged in the orthogonal Cartesian directions. Amorphous silica matrix without defects represented by a cell with $N=648$ atoms corresponding to an edge of the box $L=2.14$ nm (yellow colour) compared to a cell with $N=192$ atoms corresponding to an edge of the box $L=1.42$ nm (dark green colour) and to a cell with $N=72$ atoms corresponding to an edge box of $L=1.03$ nm (red colour). These calculations are performed with a PBE exchange-correlation functional.}
\end{figure}

In Figure \ref{fig:figureSI4} we show instead the imaginary part of the dielectric tensor for three different models of silicon dioxide corresponding to three different sizes of the boxes (yellow colour: $L=2.14$ nm, $N$=648 atoms; green colour: $L=1.42$ nm, $N$=192 atoms; red colour: $L=1.03$ nm, $N$=72 atoms). We note that the spectrum in red is characterised by a spurious peak between 7 and 8 eV, which is not present when larger boxes are used. Conversely, while the details of the main peak around 10 eV are slightly different due to the different structures, the band gap of the two systems having $L=1.42$ nm and $L=2.14$ nm is the same. This shows us that a box size of $L=1.42$ nm is sufficient to avoid spurious interactions between the periodic images when analysing the optical properties.\\
\begin{figure}[hbt!]
   \centering
\includegraphics[width=0.8\textwidth]{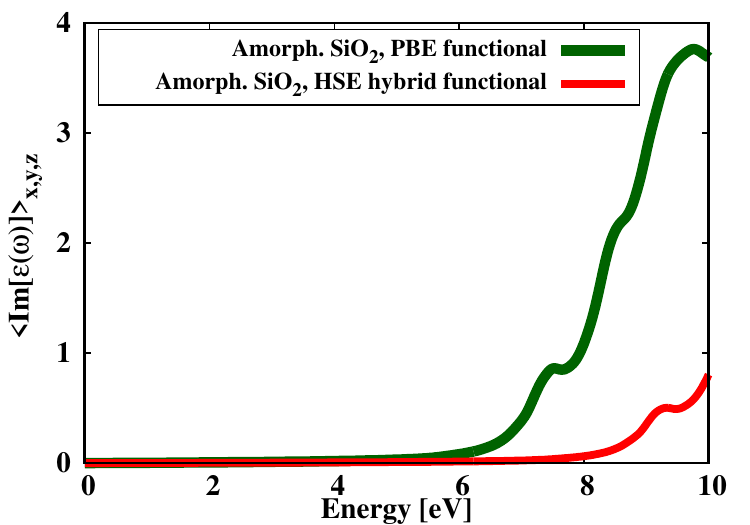}
\caption{\label{fig:figureSI6} Effect of the DFT functional: average of the imaginary part of the dielectric tensor along the orthogonal Cartesian directions. Amorphous silica matrix using a PBE exchange-correlation functional (dark green colour) compared to a hybrid HSE06 functional (red colour).}
\end{figure}
\indent In Figure \ref{fig:figureSI6} we report instead on the investigation of two different DFT functionals in the evaluation of the imaginary part of the dielectric tensor. A hybrid functional such as HSE06 reproduces the bandgap of amorphous silica much more accurately than the pure PBE exchange-correlation functional.

\section*{Fitting of Thermal Tails in Time-of-Flight Spectra}

To quantitatively investigate the thermal effects observed in the La-APT analyses in Fig. 4 of the article, we fitted the thermal tails of the Si$^{2+}$ peak in the time-of-flight (ToF) spectra using equation (2) from the main text.



The fits were performed locally within a $\sim$100~ns window after the Si$^{2+}$ peak maximum, which was selected to avoid interference from overlapping tails of subsequent peaks. Both ToF spectra were normalised to unity at the Si$^{2+}$ peak and temporally aligned such that the onset of the thermal decay ($t_0$) overlapped. The initial estimates and boundaries for the fitting parameters were physically motivated: $Q$ was varied around 0.1~eV based on the expectations of CSR-derived electric fields; $\Delta T_{\text{max}}$ ranged from 65~K to 250~K and $\tau$ was constrained between 0.5 and 10~ns, depending on the typical values observed in the literature \cite{vella2011field,vella2013interaction}.

The fitting was performed by iterative parameter adjustment using OriginLab, aiming for a qualitative visual agreement between data and model. No automated statistical optimisation was performed; the fits are therefore considered illustrative. For the fit in Figure 4b of the main text, the extracted values of $Q_{\text{deep-UV}} = 0.12$~eV and $Q_{\text{UV}} = 0.06$~eV reflect the expectation that a higher electric field of 21.5~V/nm was used for UV analyses vs. 19.7~V/nm for deep-UV analyses. The values of 65K and 250 K for $T_{\mathrm{max}}$
were determined for the UV and deep-UV analyses, respectively.

\section*{Evolution of Thermal Behavior During LEAP Analysis}

\begin{figure}[h]
\centering
\includegraphics[width=0.8\textwidth]{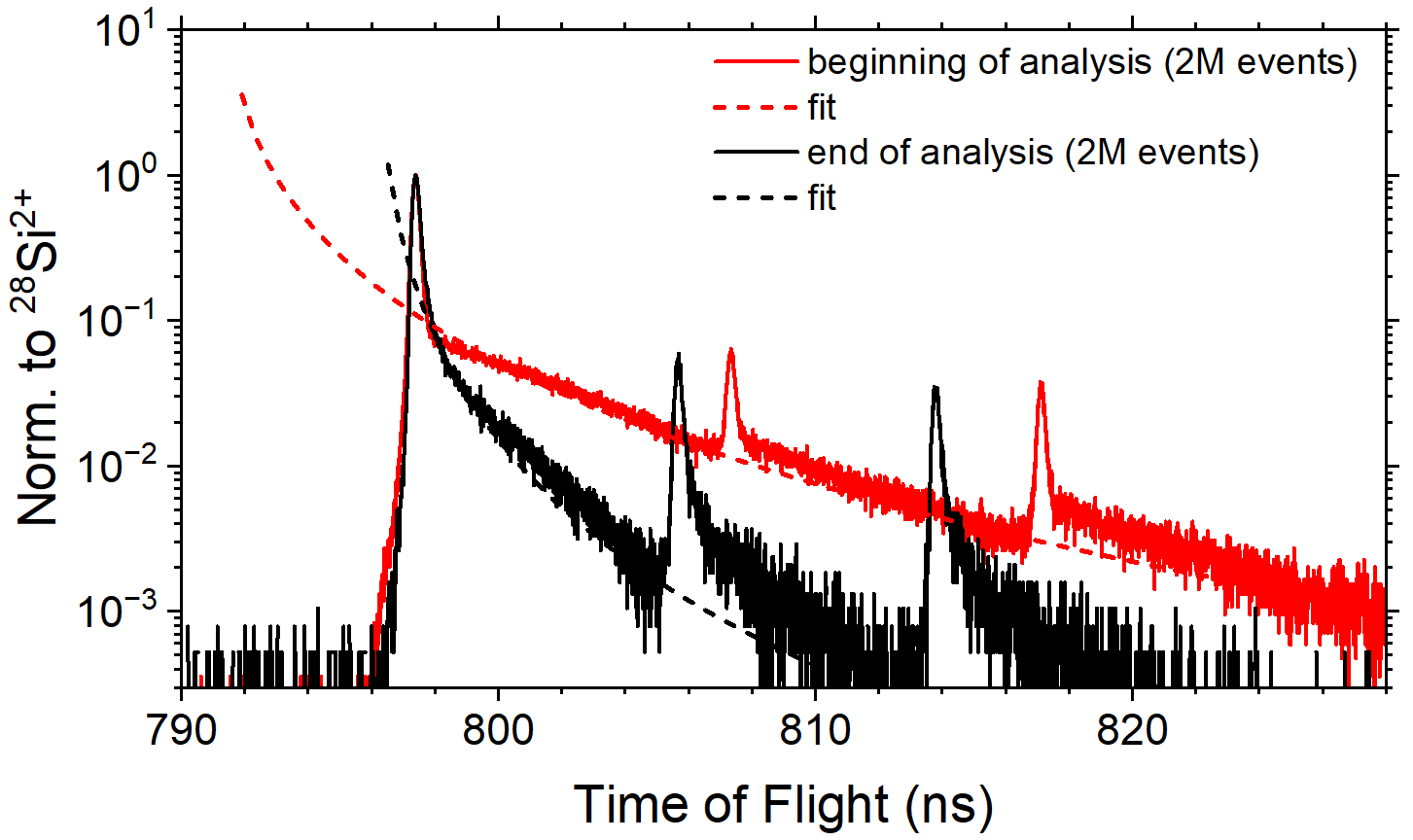}\caption{Time-of-flight (ToF) spectra of the Si$^{2+}$ peak from the beginning (red line) and end (black line) of the LEAP analysis. The first 2 million events (black curve) and the last 2 million events (red curve) were extracted from a complete data set of $\sim$16 million ions. Both spectra were normalised to unity at the peak maximum and aligned so that the onset of thermal decay ($t_0$) overlaps. The curves were fitted using equation (2) from the main text. The optimal fitting parameters for the early phase were $Q = 0.14$~eV, $\Delta T_{\text{max}} = 250$~K and $\tau = 6$~ns; for the late phase were $Q = 0.13$~eV, $\Delta T_{\text{max}} = 250$~K and $\tau = 1.5$~ns.}\label{SI_ToF.PNG}
\end{figure}

During the analysis with deep-UV light, the shape of the specimens changed, which is reflected in the change in their thermal response. Therefore, the $\sim$16 million ion dataset was split into two subsets: the first and the last 2 million events. The respective ToF spectra of the Si$^{2+}$ peak were normalised and aligned as described above and then fitted separately. The data are shown in Fig. \ref{SI_ToF.PNG}. 
For the early phase of the analysis we obtained

\begin{itemize}
 \item $Q = 0.14$~eV, $\Delta T_{\text{max}} = 250$~K, $\tau = 6$~ns
\end{itemize}

For the late phase:

\begin{itemize}
    \item $Q = 0.13$~eV, $\Delta T_{\text{max}} = 250$~K, $\tau = 1.5$~ns
\end{itemize}

The significant reduction in $\tau$ indicates an increase in the cone angle during the course of the analysis, which is confirmed by the SEM image. $T_{\mathrm{max}}$ was kept constant for both fits because the illumination conditions are constant during the analysis and the value of $Q$ is also almost constant.

\begin{figure}[htb!]
   \centering
\includegraphics[width=0.7\textwidth]{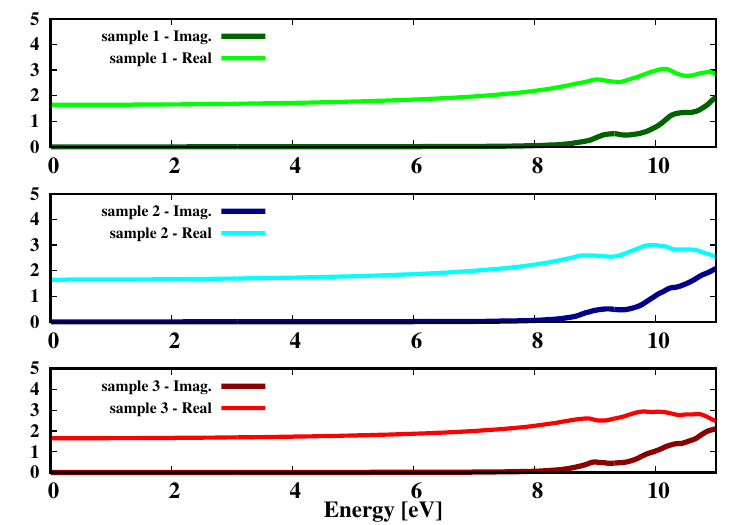}\caption{\label{fig:figureSI11} Real and imaginary part of the dielectric function for the three silicon dioxide samples in Figure 7a of the main text, calculated with the exchange-correlation functional HSE06.}
\end{figure}

\begin{figure}[htb!]
   \centering
\includegraphics[width=0.8\textwidth]{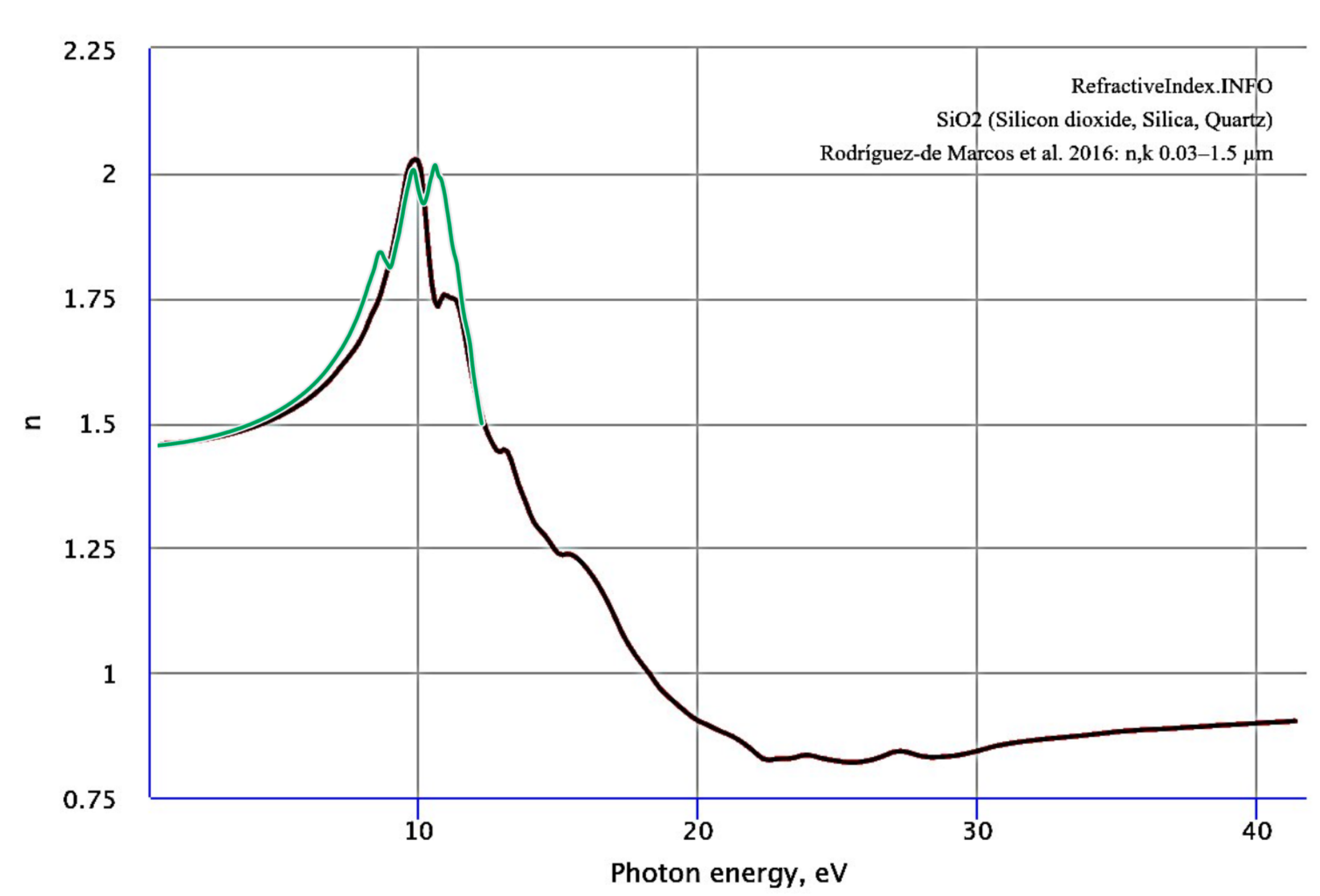}\caption{\label{fig:figureSI10} Refractive index of the silica sample 1 (green line), shown in Figure 7a of the main text, calculated with the exchange-correlation functional HSE06 in comparison to experimental data (black line).}
\end{figure}

\section*{Dielectric properties}

In Figure \ref{fig:figureSI11} we report the real and imaginary parts of the dielectric function for the three silicon dioxide samples of Figure 7a of the main text.\\ 
\indent
In Figure \ref{fig:figureSI10} we show the refractive index of the silica sample 1 of Figure 7a of the main text.

\FloatBarrier

\end{suppinfo}


\providecommand{\latin}[1]{#1}
\makeatletter
\providecommand{\doi}
  {\begingroup\let\do\@makeother\dospecials
  \catcode`\{=1 \catcode`\}=2 \doi@aux}
\providecommand{\doi@aux}[1]{\endgroup\texttt{#1}}
\makeatother
\providecommand*\mcitethebibliography{\thebibliography}
\csname @ifundefined\endcsname{endmcitethebibliography}  {\let\endmcitethebibliography\endthebibliography}{}

\end{document}